\documentclass[secnumarabic, twocolumn, amssymb, superscriptaddress, aps, prl,longbibliography]{revtex4-2}
\usepackage{graphicx}
\usepackage{amsmath}
\usepackage{upgreek}
\usepackage{color}
\usepackage{setspace}
\usepackage{hyperref}
\hypersetup{colorlinks,breaklinks,
            urlcolor=[rgb]{0,0,0.64},
            linkcolor=[rgb]{0,0,0.64},
            citecolor=[rgb]{0,0,0.64}}
\usepackage{nccmath}
\usepackage{amssymb}
\usepackage{bm}
\setcitestyle{super}
\usepackage{multirow}
\usepackage{makecell}
\usepackage[T2A]{fontenc}
\usepackage[utf8]{inputenc}

\newcommand{\KLLP}{Key Laboratory for Laser Plasmas (Ministry of Education), School of Physics and Astronomy, Shanghai Jiao Tong University, Shanghai 200240, China.}
\newcommand{\CICIFSA}{Collaborative Innovation Center of IFSA (CICIFSA), Shanghai Jiao Tong University, Shanghai 200240, China.}
\newcommand{\Cal}{Department of Chemistry, University of California at Berkeley, Berkeley, California 94720, USA.}
\newcommand{\LBNL}{Materials Sciences Division, Lawrence Berkeley National Laboratory, Berkeley, California 94720, USA.}
\newcommand{\UCLA}{Department of Physics and Astronomy, University of California at Los Angeles, Los Angeles, California 90095, USA.}
\newcommand{\CAS}{Beijing National Laboratory for Condensed Matter Physics, Institute of Physics, Chinese Academy of Sciences, Beijing 100190, China.}
\newcommand{\KLASQC}{Key Laboratory of Artificial Structures and Quantum Control (Ministry of Education), Shenyang National Laboratory for Materials Science, School of Physics and Astronomy, Shanghai Jiao Tong University, Shanghai 200240, China.}
\newcommand{\Tsinghua}{State Key Laboratory of Low Dimensional Quantum Physics, Department of Physics, Tsinghua University, Beijing 100084, China.}
\newcommand{\SHTech}{School of Physical Science and Technology, ShanghaiTech University, Shanghai 201210, China.}
\newcommand{\SHTechTP}{ShanghaiTech Laboratory for Topological Physics, Shanghai 201210, China.}
\newcommand{\TDLI}{Tsung-Dao Lee Institute, Shanghai Jiao Tong University, Shanghai 200240, China.}
\newcommand{\ZIAS}{Zhangjiang Institute for Advanced Study, Shanghai Jiao Tong University, Shanghai 200240, China.}

\begin{document}

\title{Light-induced dimension crossover in 1\textit{T}-TiSe$_2$ dictated by excitonic correlations}

\author{Yun~Cheng}
\thanks{These authors contributed equally: Y.C. and A.Z.}
\affiliation{\KLLP}
\affiliation{\CICIFSA}

\author{Alfred~Zong}
\thanks{These authors contributed equally: Y.C. and A.Z.}
\affiliation{\Cal}
\affiliation{\LBNL}

\author{Jun~Li}
\affiliation{\CAS}

\author{Wei~Xia}
\affiliation{\SHTech}
\affiliation{\SHTechTP}

\author{Shaofeng~Duan}
\affiliation{\KLASQC}

\author{Wenxuan~Zhao}
\affiliation{\Tsinghua}

\author{Yidian~Li}
\affiliation{\Tsinghua}

\author{Fengfeng~Qi}
\affiliation{\KLLP}
\affiliation{\CICIFSA}

\author{Jun~Wu}
\affiliation{\KLLP}
\affiliation{\CICIFSA}

\author{Lingrong~Zhao}
\affiliation{\KLLP}
\affiliation{\CICIFSA}

\author{Pengfei~Zhu}
\affiliation{\KLLP}
\affiliation{\CICIFSA}

\author{Xiao~Zou}
\affiliation{\KLLP}
\affiliation{\CICIFSA}

\author{Tao~Jiang}
\affiliation{\KLLP}
\affiliation{\CICIFSA}

\author{Yanfeng~Guo}
\affiliation{\SHTech}

\author{Lexian~Yang}
\affiliation{\Tsinghua}

\author{Dong~Qian}
\affiliation{\KLASQC}

\author{Wentao~Zhang}
\affiliation{\KLASQC}

\author{Anshul~Kogar}
\email{anshulkogar@physics.ucla.edu}
\affiliation{\UCLA}

\author{Michael~W.~Zuerch}
\email{mwz@berkeley.edu}
\affiliation{\Cal}
\affiliation{\LBNL}

\author{Dao~Xiang}
\email{dxiang@sjtu.edu.cn}
\affiliation{\KLLP}
\affiliation{\CICIFSA}
\affiliation{\TDLI}
\affiliation{\ZIAS}

\author{Jie~Zhang}
\email{jzhang1@sjtu.edu.cn}
\affiliation{\KLLP}
\affiliation{\CICIFSA}
\affiliation{\TDLI}

\date{\today}

\begin{abstract}
In low-dimensional systems with strong electronic correlations, the application of an ultrashort laser pulse often yields novel phases that are otherwise inaccessible. The central challenge in understanding such phenomena is to determine how dimensionality and many-body correlations together govern the pathway of a non-adiabatic transition. To this end, we examine a layered compound, 1\textit{T}-TiSe$_\text{2}$, whose three-dimensional charge-density-wave (3D CDW) state also features exciton condensation due to strong electron-hole interactions. We find that photoexcitation suppresses the equilibrium 3D CDW while creating a nonequilibrium 2D CDW. Remarkably, the dimension reduction does not occur unless bound electron-hole pairs are broken. This relation suggests that excitonic correlations maintain the out-of-plane CDW coherence, settling a long-standing debate over their role in the CDW transition. Our findings demonstrate how optical manipulation of electronic interaction enables one to control the dimensionality of a broken-symmetry order, paving the way for realizing other emergent states in strongly correlated systems.
\end{abstract}

\maketitle

A symmetry-breaking phase transition in 1D or 2D is qualitatively distinct from its 3D counterpart. Strictly speaking, a 1D chain with short-range interactions cannot undergo a thermodynamic transition at finite temperatures \cite{VanHove1950}. With short-range interactions in a 2D plane, a transition can occur through the binding or dissociation of vortex-antivortex pairs \cite{Kosterlitz1973}. In real materials, however, a more complex situation arises due to nonzero couplings between chains or planes. Although long-range order is possible, significant order parameter fluctuations prevent a mean-field treatment of the transition.

\begin{figure*}[htb!]
	\includegraphics[width=0.85\textwidth]{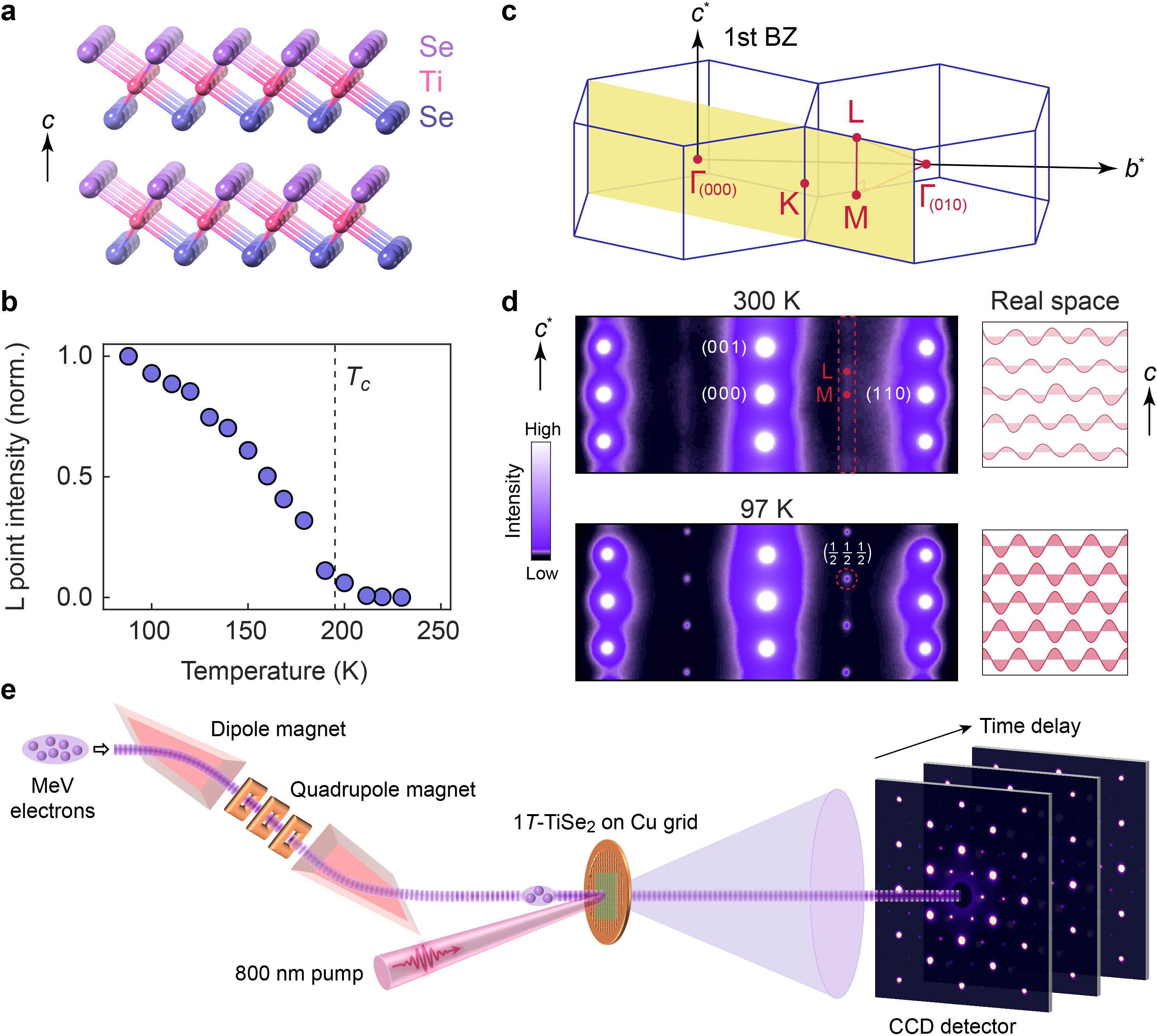}
	\caption{\textbf{Equilibrium charge density wave transition in 1$\bm{T}$-TiSe$_\text{2}$}. \textbf{a},~Crystal structure of 1$T$-TiSe$_2$ in the non-CDW state. Layers are bonded by van der Waals forces and each Ti atom is octahedrally coordinated with six Se atoms, where upper and lower Se atoms are colored differently. \textbf{b},~Temperature-dependent intensity evolution of the $\left(\frac{1}{2}~\frac{1}{2}~\frac{1}{2}\right)$ CDW peak, which is located at the L point in the Brillouin zone (BZ). Intensity is normalized by the value at 88~K. \textbf{c},~Brillouin zones of 1$T$-TiSe$_2$ with high symmetry points labeled. The yellow plane corresponds to the diffraction plane in \textbf{d}. \textbf{d},~Static electron diffraction patterns in the $[1~\bar{1}~0]$ zone axis above and below the CDW transition temperature, $T_c\approx195$~K. Diffuse streaks along the M--L line at 300~K (dashed rectangle) is transformed into sharp superlattice peaks at the L point at 97~K (dashed circle). Schematics on the right show the real space configurations of CDWs in different layers. \textbf{e},~Schematic of the ultrafast electron diffraction setup, where a double-bend achromatic lens consisting of a pair of dipole magnets and three quadrupole magnets are used to compress the electron pulse and to reduce the timing jitter (see Methods).}
\label{fig:1}
\end{figure*}

When a system is characterized by both reduced dimensionality and strong electronic correlations, large fluctuations coupled with several competing energy scales provide a fruitful ground for realizing exotic states of matter. This intuition in thermal equilibrium can be leveraged to access new phases using an ultrafast laser pulse, which serves to both modify the Coulomb interaction via carrier screening and to enhance fluctuations when the excited carriers relax \cite{Sun2020}. For example, shining a single near-infrared pulse onto the Mott insulating state of 1$T$-TaS$_2$ produces a ``hidden'' metallic phase \cite{Stojchevska2014}, and illuminating a PbTiO$_3$/SrTiO$_3$ heterostructure film with an optical pulse leads to a metastable supercrystal \cite{Stoica2019}. Whereas carrier screening can be well captured by theoretical models to account for such light-induced phenomena \cite{Karpov2018,Stoica2019}, out-of-equilibrium fluctuations in low-dimensional systems remain poorly understood due to a lack of experimental probes available to quantify fluctuations along different spatial dimensions. In this article, we study a quasi-2D material with strong electron-hole correlations, 1$T$-TiSe$_2$, which also hosts a commensurate CDW. Following femtosecond laser excitation, we observed large, anisotropic CDW fluctuations using MeV ultrafast electron diffraction (see Methods and Supplementary Note~1). Compared to other ultrafast probes, this technique is capable of detecting extremely weak diffuse signals in a momentum-resolved manner,  unveiling a 3D-to-2D CDW crossover. We show that the transient appearance of the 2D density wave is controlled by the screening of exciton correlations, providing a unique perspective on the interplay between order parameter fluctuation and electronic correlation in low-dimensional materials.

\newpage
\begin{figure*}[htb!]
\centering
    \includegraphics[width=1\textwidth]{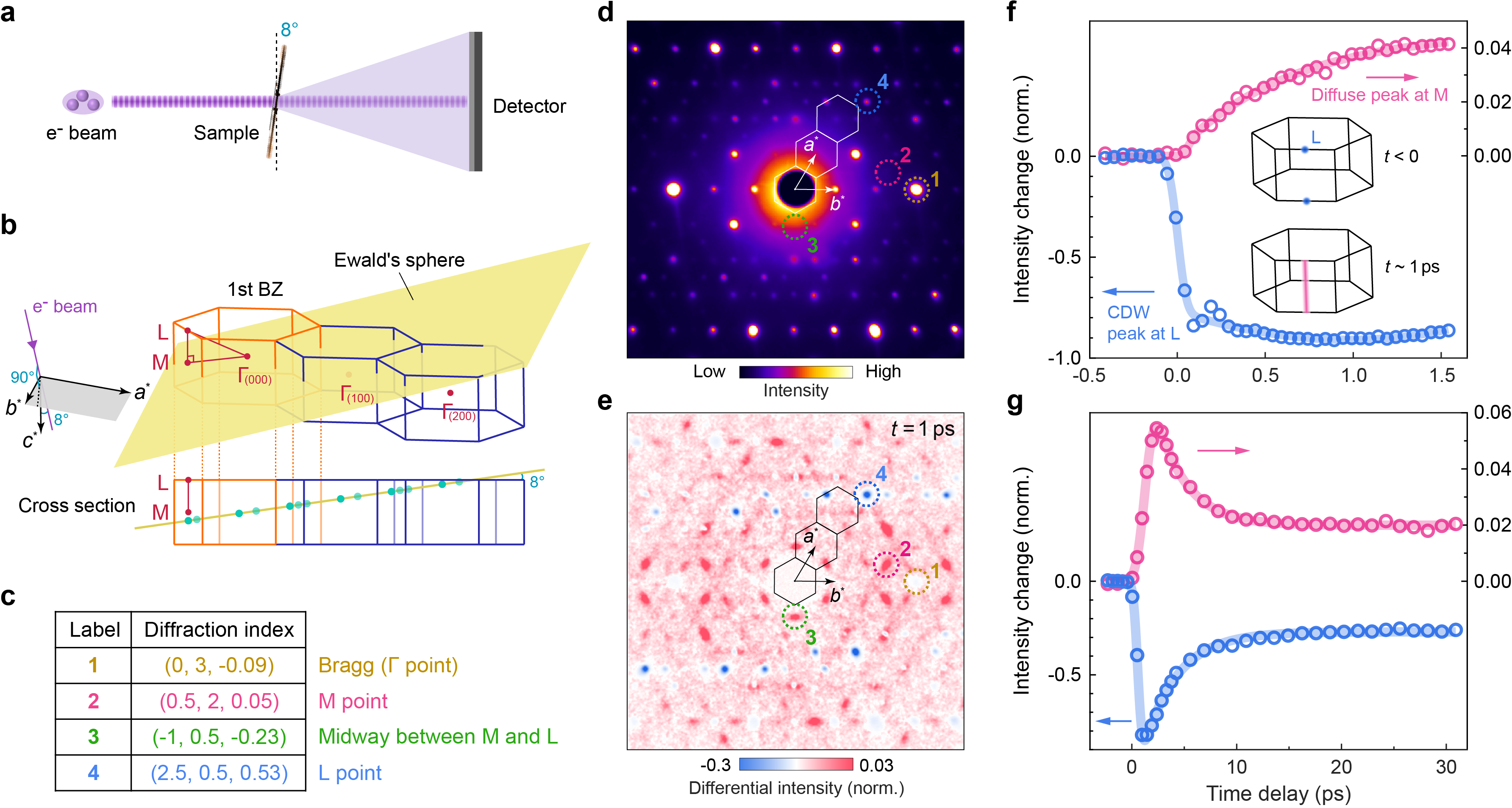}
    \caption{\textbf{Light-induced dimension crossover from a 3D to 2D CDW.} \textbf{a},~Schematic of electron diffraction geometry for data presented in panels \textbf{d}--\textbf{g}, where an ultrathin sample cleaved in the (0~0~1) plane is rotated by 8$^\circ$ relative to the incident electron beam. \textbf{b},~Illustration of the intersection between the reciprocal lattice and the Ewald's sphere, which is locally approximated by a plane that corresponds to the diffraction pattern in \textbf{d} and \textbf{e}. Green dots in the cross-sectional view mark the intersection points between the L--M--L line and the Ewald's sphere. \textbf{c},~Miller indices for four color-coded points in the diffraction images. \textbf{d},~Static electron diffraction pattern taken at 88~K in the tilted geometry with zone axis equal to $(\overline{36}~5~164)$. The three hexagons correspond to the projection of the three Brillouin zones in \textbf{b} onto the Ewald's sphere. \textbf{e},~Differential diffraction pattern at 1~ps after photoexcitation by an 800-nm, 30-fs pulse. \textbf{f},~Time evolution of the CDW peak at the L point (peak~4, blue) and the diffuse peak at the M point (peak~2, red). Intensity values are normalized by their respective averages before photoexcitation. See Fig.~S5b for traces of additional peaks. Inset illustrates the photoinduced change in the CDW dimensionality, sketched in the reciprocal space. \textbf{g},~The same as \textbf{f} but plotted for extended pump-probe time delay, showing the partial recovery to a quasi-equilibrium plateau. In \textbf{f} and \textbf{g}, error bars of intensity change are smaller than the marker size; solid curves are fits to an error function multiplied by an exponential function. The incident fluence for data presented in \textbf{e}--\textbf{g} is 560~$\upmu$J/cm$^2$.} 
\label{fig:2}
\end{figure*}

1$T$-TiSe$_2$ has a layered structure where adjacent planes are coupled by van der Waals forces (Fig.~\ref{fig:1}a). A 3D $(2\times2\times2)$ CDW forms below $T_c\approx 195$~K, evidenced by the development of the corresponding superlattice peak (Fig.~\ref{fig:1}b). In monolayer samples or in bulk crystals above $T_c$, a 2D version of the CDW was also reported \cite{Chen2015,Chen2016,Watson2020}, making it an ideal platform to investigate the morphing dimensions of the symmetry-breaking order. The CDW transition in 1$T$-TiSe$_2$ is distinguished by concurrent exciton condensation due to the large binding energy between holes in the Se 4$p$ band and electrons in the Ti 3$d$ band \cite{Kogar2017}. However, the exact contribution by the excitonic interaction to the CDW transition has not been clearly identified despite close scrutiny over four decades \cite{Hughes1977,Suzuki1985,Kidd2002,Cercellier2007,Calandra2011,Porer2014,Kogar2017,Burian2021}. Here, by correlating the ultrafast response of bound electron-hole pairs to the CDW fluctuations, we show that excitonic interactions are primarily responsible for the out-of-plane coherence of the CDW. Our work hence offers important insight into the role of electronic correlations in driving the symmetry-breaking transition.

We first characterize the equilibrium 3D CDW using a transmission electron microscope (TEM). The unit cell doubling in the out-of-plane axis leads to superlattice peaks that, to the first order, only appear at the L point instead of the M point in reciprocal space (Fig.~\ref{fig:1}c), which are verified in our diffraction pattern in the [$1~\bar{1}~0$] zone axis. A cut in this plane covers both L and M momenta values (Fig.~\ref{fig:1}d). As the sample is heated above $T_c$, the sharp peak at L morphs into an elongated streak along the M--L line. From the Fourier transform relation, these streaks correspond to short-range CDWs within each plane that lack any out-of-plane order. Our temperature-dependent measurements establish that diffraction intensities at different locations along the M--L line are a sensitive probe of the CDW dimensionality: A sharp peak solely at L represents a 3D $(2\times2\times2)$ CDW while diffuse intensity from M to L indicates a 2D $(2\times2)$ order without interplane coherence (see schematics in Fig.~\ref{fig:1}d).

We exploit these signatures of the CDW dimension to study its evolution after photoexcitation. Unlike a standard TEM, the time-resolved setup cannot focus electrons very tightly to minimize temporal pulse broadening. Hence, ultrathin samples with large lateral dimensions are required \cite{Bie2021}, obtained by exfoliating a bulk crystal that naturally cleaves in the (0~0~1) plane (see Methods). To measure the M--L line, we tilt the exfoliated flake by 8$^\circ$ relative to the incoming electron beam (Fig.~\ref{fig:2}a). In this way, the Ewald sphere intersects the M--L line at different locations in different Brillouin zones, enabling us to sample the intensity distribution from M to L (green dots in Fig.~\ref{fig:2}b). The static diffraction at 88~K in this tilted geometry is shown in Fig.~\ref{fig:2}d, where several momenta between M and L are identified (Fig.~\ref{fig:2}c). As expected from a 3D long-range CDW, superlattice peaks only appear at L while absent at other locations along the M--L cut, in excellent agreement with the simulated pattern (Fig.~S2 and Supplementary Note~2).

The photoinduced change at 88~K following an 800-nm, 30-fs pulse is presented in the differential intensity plot in Fig.~\ref{fig:2}e, where we subtract the diffraction pattern before photoexcitation from the one at 1~ps delay. Here, we focus on the fast CDW dynamics along the M--L line, during which Debye-Waller heating of the lattice has negligible contribution (see Supplementary Note~3). Along the M--L cut, a transient intensity decrease is seen at L (location~4) while enhancements are observed elsewhere (locations~2 and 3), leading to clear diffuse peaks shown in red. To understand the contrasting changes, we examine their timescales by plotting the intensity evolution at L and M (Figs.~\ref{fig:2}f and S5b). There is a clear separation of the initial response times: The intensity drop at L is completed within 200~fs while the peak growth at M takes more than 1~ps. The time trace of the CDW peak at L also displays a strongly damped oscillation (blue circles in Figs.~\ref{fig:2}f and S5b) that is commonly observed in the ultrafast melting of a charge-ordered state \cite{Huber2014,Moore2016,Wall2018}, whereas no such oscillation is detected in the diffuse peak at M (red circles in Fig.~\ref{fig:2}f). These differences between the L and M points are reminiscent of the timescale dichotomy in the transient creation of a competing CDW in LaTe$_3$ \cite{Zong2021} and during the hidden state formation in 1$T$-TaS$_2$ \cite{Ravnik2018}, suggesting that the destruction of the 3D CDW is driven by a coherent displacive mechanism while 2D CDW fluctuations develop in a spatially inhomogeneous fashion.

\begin{figure}[htb!]
    \includegraphics[width=0.373\textwidth]{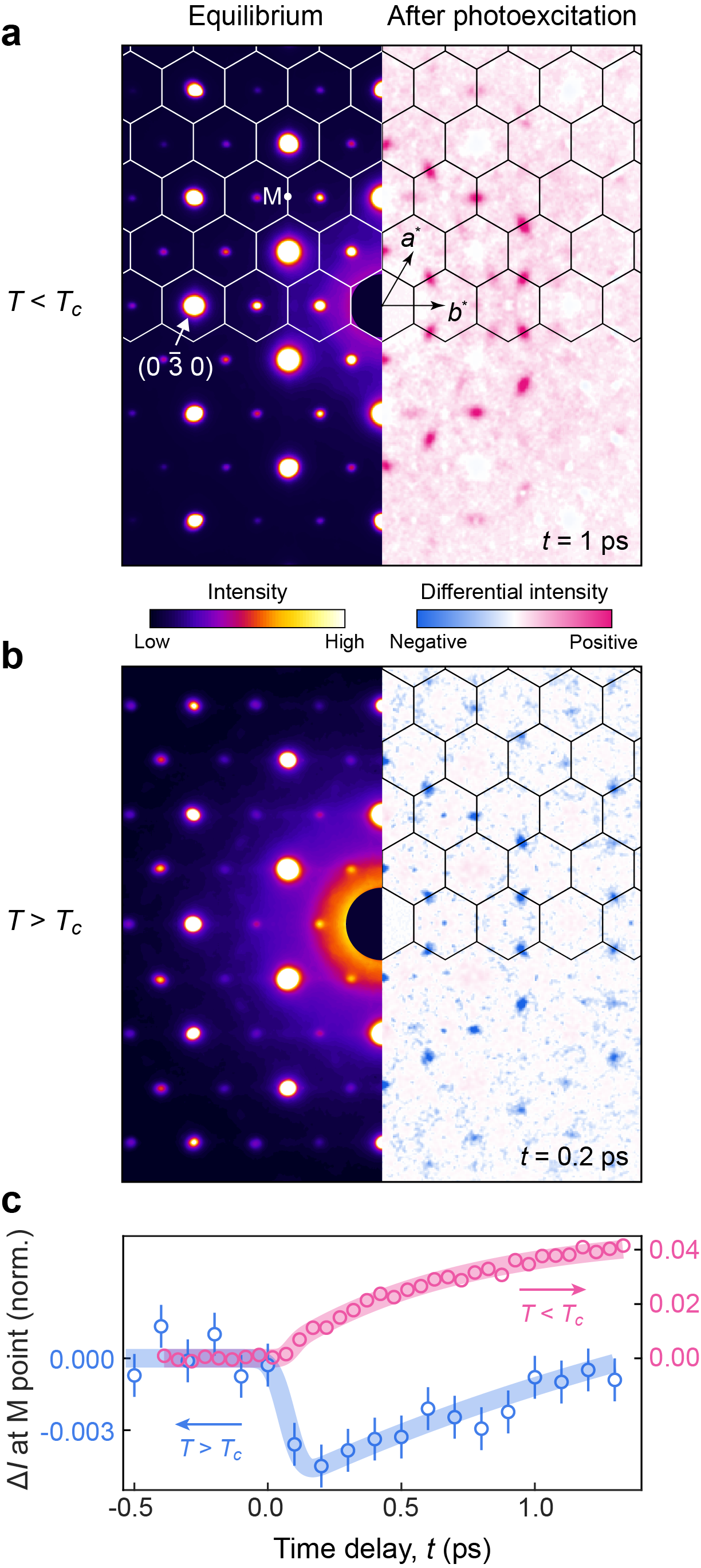}
    \caption{\textbf{Distinct photoinduced changes of 2D CDW below and above $\bm{T}_{\bm{c}}$.} \textbf{a},\textbf{b},~\textit{Left}: static electron diffraction patterns along the [0~0~1] zone axis taken at 88~K (\textbf{a}) and 295~K (\textbf{b}). Hexagons denote the 2D projection of the Brillouin zones. \textit{Right}: photoinduced change in the diffraction intensity at 1~ps pump-probe delay at 88~K (\textbf{a}) and at 0.2~ps delay at 250~K (\textbf{b}). The different time delays are chosen to reflect the distinct timescales of the intensity change at different temperatures (see panel \textbf{c}). All diffraction images were symmetrized for enhanced statistics. The equilibrium patterns share the same color scale while that of the differential patterns is individually adjusted to highlight the feature at the M point. See Fig.~S8 for a side-by-side comparison of differential patterns at two time delays with the same color scale. \textbf{c},~Temporal evolutions of intensity change at the M point at 88~K (red) and 250~K (blue). Intensities are normalized to respective pre-excitation values. Error bars, if larger than the marker size, represent the standard deviation of intensity values prior to excitation. The incident fluence for all panels is 560~$\upmu$J/cm$^2$.}
\label{fig:3}
\end{figure}

\begin{figure*}[htb!]
    \includegraphics[width=0.7\textwidth]{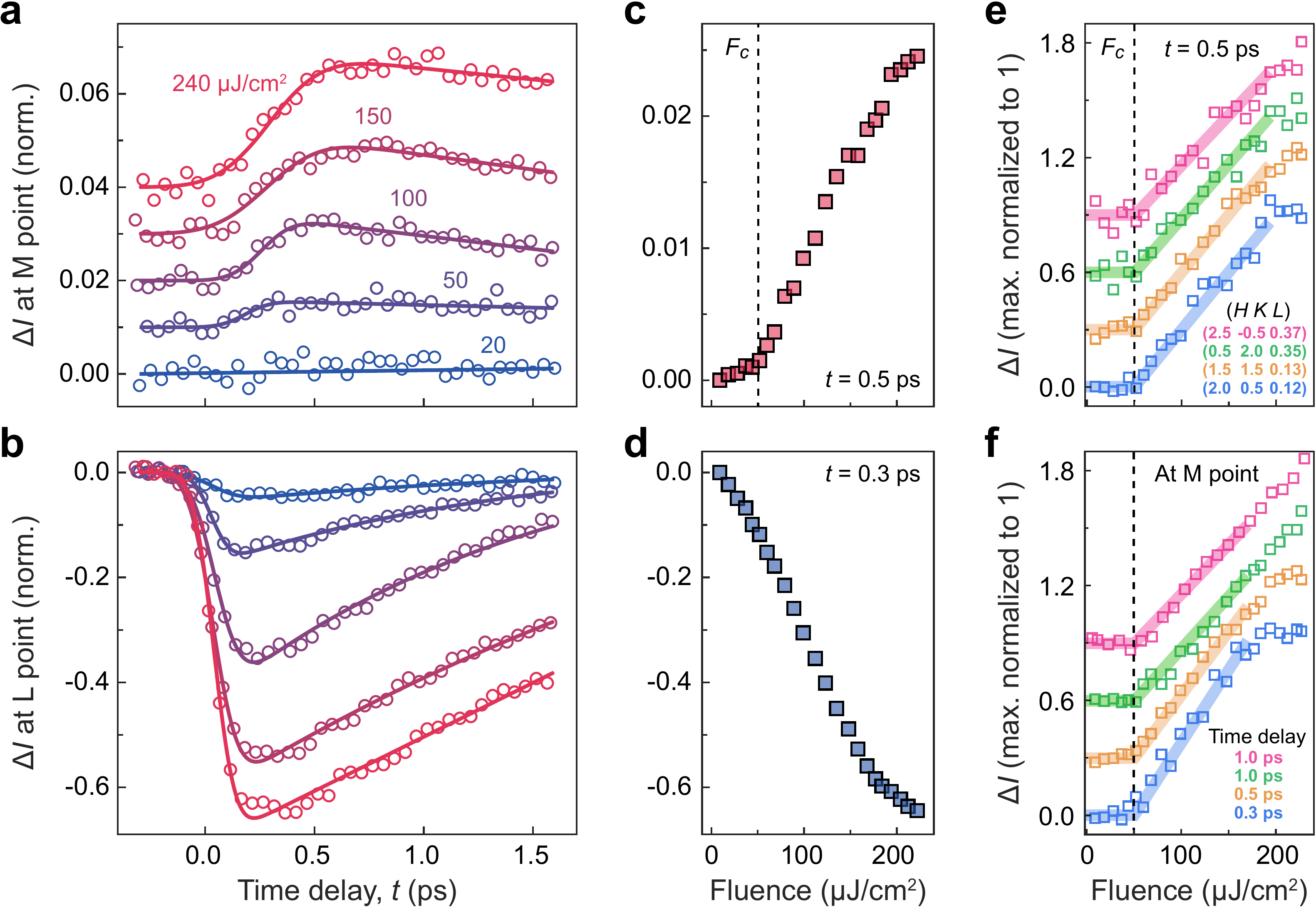}
    \caption{\textbf{Fluence-dependent dimension crossover of the CDW.} \textbf{a},\textbf{b},~Time evolution of changes in integrated intensities ($\Delta I$) at M and L points, measured at 88~K. Intensity values are normalized by the averages before photoexcitation. Incident fluences are labeled and color-coded. Solid curves are fits to an error function multiplied by an exponential relaxation and serve as a guide to the eye. For the M point, multiple diffuse peaks are used for enhanced signal-to-noise ratio. For the L point, superlattice peak $\left(2~\frac{1}{2}~\frac{1}{2}\right)$ is used. \textbf{c},\textbf{d},~Fluence-dependent changes in intensity at the M point at 0.5~ps time delay (\textbf{c}) and at the L point at 0.3~ps time delay (\textbf{d}). The vertical dashed line marks the critical fluence, $F_c$, above which the photoinduced intensity at the M point starts to rapidly increase. A separate critical fluence is observed in the L point intensity at long time delay (see Fig.~S4 and Supplementary Note~4). \textbf{e},~Fluence-dependent intensity change at various momenta along the M--L cut, measured at a fixed time delay of 0.5~ps. \textbf{f},~Fluence-dependent intensity change at various pump-probe delays, measured at the M point. Red data points at 1.0~ps were taken from a separate sample compared to the other data points. In \textbf{e} and \textbf{f}, traces are normalized so that the maximum change is 1; the traces are also vertically shifted by 0.3 for clarity. Curves are guides to the eye.}
\label{fig:4}
\end{figure*}

Taken together, the intensity evolution along the M--L line reveals a dimension crossover from a 3D to 2D CDW, signified by a transformation of the sharp peak at L into a continuous streak stretching the entire M--L cut (inset in Fig.~\ref{fig:2}f). The extent of the dimension crossover can be inferred from a comparison of the absolute intensity among different points along the M--L line (Fig.~S6 and Supplementary Note~5): For the fluence corresponding to Fig.~\ref{fig:2}, photoexcitation yields a transient 2D CDW with nearly no phase coherence between neighboring layers. At long pump-probe delay, the partial recovery to the original 3D CDW is marked by the same relaxation timescale in both M and L peaks when some out-of-plane coherence is re-gained (Fig.~\ref{fig:2}g). After 10~ps, the system enters a quasi-equilibrium state, which takes more than $\sim 2$~ns to relax to the pre-excitation configuration (Fig.~S4b and Supplementary Note~4). The slow recovery of the long-range 3D CDW may be attributed to the presence of photoinduced defects \cite{Vogelgesang2018,Zong2019}, which disrupt the CDW phase coherence and account for the persistent broadening of the superlattice peak (Fig.~S4c).

The light-induced 2D CDW is only observed below the equilibrium $T_c$. Above $T_c$, photoexcitation only serves to transiently reduce the equilibrium fluctuations. This contrast is best illustrated by the differential diffraction patterns along the [0~0~1] zone axis, measured below and above $T_c$ under the same photoexcitation conditions (Figs.~\ref{fig:3} and S8). At 250~K, the intensity reduction at M (Figs.~\ref{fig:3}b and S8b) forms a hexagonal pattern with intensities determined by the geometric structure factor \cite{Otto2021}. This intensity decrease indicates a rapid suppression of the amplitude of the 2D density waves within 0.2~ps (Fig.~\ref{fig:3}c and Supplementary Note~6). Its stark contrast to the slower intensity enhancement below $T_c$ (Figs.~\ref{fig:3}a,c and S8a,c) suggests that the photoinduced destruction of a 3D CDW in a quasi-2D environment is manifested by a dimension crossover that preserves the in-plane order.

\begin{table*}
\caption{\label{tab:table1}Energy threshold for melting the exciton order based on different observables in various pump-probe techniques, expressed in terms of absorbed energy per normal state unit cell (u.c.). See Supplementary Note~7 for the calculation of the absorbed energy density.}
\vspace{8pt}
\small
\begin{tabular}{c|c|c|c}
\hline\hline
\textbf{~Time-resolved probe~}  & \textbf{Observable} & \makecell{\textbf{~Threshold~}\\ \textbf{(meV/u.c.)}} & \textbf{Reference}\\
\hline\hline
Electron diffraction & Diffuse 2D CDW peak & 1.8 & This work \\
\hline
\multirow{2}{*}{\makecell{Terahertz\\spectroscopy}} & Plasmon resonance & \multirow{2}{*}{1.8} & \multirow{2}{*}{Porer \textit{et al.}\cite{Porer2014}} \\
\cline{2-2}
  & \multirow{2}{*}{Coherent oscillation} & & \\
\cline{1-1}\cline{3-4}
Optical reflectivity &  &  \multirow{2}{*}{2.6} & \multirow{2}{*}{~Hedayat \textit{et al.}\cite{Hedayat2019}~ } \\
\cline{1-2}
\multirow{3}{*}{\makecell{Angle-resolved \\ photoemission}} & \multirow{2}{*}{Se-4$p$ valence band} & & \\
\cline{3-4}
   &  & \multirow{2}{*}{3.1} & \multirow{2}{*}{Duan \textit{et al.} \cite{Duan2021} } \\
\cline{2-2}
   & ~Excited carrier relaxation~ & & \\
\hline
\makecell{Resonant x-ray\\diffraction} & \makecell{Space-group forbidden\\CDW peak} & 1.35 & Burian \textit{et al.} \cite{Burian2021} \\
\hline\hline
\end{tabular}
\end{table*}

One is tempted to interpret the photoinduced crossover from the perspective of the equilibrium transition, where fluctuations of the CDW order reach a maximum at $T_c$ \cite{Holt2001}. If carrier excitation were to produce similar effects to thermal heating, one would expect transient enhancement of diffuse scattering along M--L at temperatures below $T_c$ but transient suppression instead above $T_c$, consistent with Fig.~\ref{fig:3}a,b. However, such an analogy to the equilibrium transition breaks down when we investigate the fluence dependence of the nonequilibrium crossover. As the incident fluence is increased, the fast suppression and slow enhancement at L and M, respectively, become more pronounced (Fig.~\ref{fig:4}a,b). For a quantitative comparison, we track the intensity changes at M and L as a function of fluence while fixing the delay time, $t$, at which the photoinduced change is maximal during the ultrafast evolution (Fig.~\ref{fig:4}c,d). As the maximum of the intensity change at M shifts between 0.3~ps and 0.7~ps upon increasing the fluence (Fig.~\ref{fig:4}a), we use the average time delay at $t=0.5$~ps in Fig.~\ref{fig:4}c. The most striking feature in Fig.~\ref{fig:4}c is the existence of a fluence threshold at $F_c \approx 50~\upmu$J/cm$^2$, below which negligible intensity enhancement at M is observed. By contrast, no anomaly is observed across $F_c$ at the L point (Fig.~\ref{fig:4}d). To confirm the reproducibility of this fluence threshold, we plot the fluence-dependent intensity change at other pump-probe delays and other momenta along the M--L cut, measured in different samples (Fig.~\ref{fig:4}e,f). The same value of $F_c$ is observed solely in the diffuse scattering signal away from L in the M--L cut that represents the transient formation of a 2D CDW. This distinction between the M and L points suggests that below $F_c$, laser pulses weaken the amplitude of the 3D CDW but the interplane coherence is not affected. Only at fluences above $F_c$ does the dimensional crossover start to take place. 

The existence of a fluence threshold is unexpected from equilibrium considerations, where a temperature increase towards $T_c$ continuously strengthens CDW fluctuations \cite{Holt2001}. As the initial effect of photoexcitation at 1.55~eV is free carrier generation, the fluence threshold hints at a critical carrier density that controls the dimension crossover in the CDW. The observed value of $F_c$ is translated into a transient free carrier density of $\sim 4\times10^{20}~$cm$^{-3}$ (see Supplementary Note~7), which corresponds to a Thomas-Fermi screening length just short of the superlattice period both in and out-of-plane \cite{Porer2014}. Hence, Coulomb attraction between electrons and holes on the length scale of the CDW wavelength is strongly suppressed above $F_c$, marking a transient breakdown of the excitonic correlations. The link between $F_c$ and the dissociation of bound electron-hole pairs is corroborated by several time-resolved experiments that are sensitive to carrier dynamics (Table~\ref{tab:table1}) \cite{Porer2014,Hedayat2019,Duan2021,Burian2021}. In time-resolved terahertz spectroscopy, a plasmon pole in the energy loss function is significantly modified above $F_c$, suggesting that photoexcitation has transiently melted the exciton condensate \cite{Porer2014}. This conclusion is cross-validated by femtosecond resonant x-ray diffraction. At $F_c$, there is a complete suppression of a special type of superlattice peaks \cite{Burian2021}, which are space-group forbidden and can only be observed if excitonic correlations are present. Time-resolved optical reflectivity and photoemission also demonstrate a similar threshold at $F_c$ in the valence band shift and carrier relaxation rate, both of which are sensitive to the excitonic order \cite{Hedayat2021,Duan2021}. All of these observations point towards a critical free carrier density at $F_c$, above which excitonic correlations transiently disappear.
 
Such interpretation of the fluence threshold suggests that the interlayer CDW coherence persists unless the exciton condensate is first melted due to carrier excitation and a screened potential. The chief role of exciton correlation in the CDW transition is thus to maintain the interplane coherence, thanks to the long-range nature of Coulomb coupling between Ti 3$d$ electrons and Se 4$p$ holes. Below the fluence threshold $F_c$, the effective excitonic correlation length exceeds twice the interlayer spacing, hence contributing to locking the CDW phase in the out-of-plane direction. Our analysis also offers a natural explanation for why the CDW lattice distortion persists even if excitons are destroyed by photoexcitation \cite{Porer2014}: The non-thermal separation between the lattice and excitonic orders is a manifestation of the evolving CDW dimension.

The investigation of 1$T$-TiSe$_2$ yields new insights toward the understanding of nonequilibrium phenomena that can arise in low-dimensional materials with strong electronic correlations, from light-induced superconductivity in cuprates \cite{Kaiser2017} to the insulator-metal transition in 1$T$-TaS$_2$ \cite{Stojchevska2014,Stahl2020}. Not only does our work establish dimensional crossover as the principal pathway in a non-adiabatic transition, it also demonstrates a viable route to control ordered phases with light through impulsive modification of Coulomb interactions, providing access to novel states of matter that are hidden in thermal equilibrium.

\section{Methods}
\noindent\textbf{Sample preparation.} High-quality single crystals of 1$T$-TiSe$_2$ were grown by chemical vapour transport with an iodine transport agent. Ti and Se were mixed in a molar ratio of 1:2 and placed into an alumina crucible before sealed into a quartz tube. The quartz tube was heated to 700$^\circ$C and 1$T$-TiSe$_2$ crystals were synthesized at the 650$^\circ$C zone for two weeks. 1$T$-TiSe$_2$ thin flakes were obtained by repeated exfoliation of the bulk crystal with polydimethylsiloxane films (PDMS, Gel-Pak). Flakes were pre-screened for thickness and uniformity with an optical microscope using the color contrast and further characterized by atomic force microscopy. Selected flakes were detached from PDMS in ethanol and scooped onto a 600~mesh/inch TEM grid. The resulting free-standing flake has a typical lateral dimension of $\sim 300~\upmu$m and thickness of $\sim 60$~nm.\\

\noindent\textbf{MeV ultrafast electron diffraction.} The schematic of the MeV UED beamline is shown in Fig.~\ref{fig:1}e. The 800-nm (1.55-eV), 30-fs pulses from a Ti:sapphire regenerative amplifier system operating at a repetition rate of 100~Hz (Vitara and Legend Elite Duo HE, Coherent) were split into pump and probe branches. The probe branch was frequency tripled in nonlinear crystals before illuminating a photocathode for electron pulse generation. After being accelerated by an intense radio-frequency field to relativistic velocity ($\sim 0.989c$), the electron beam went through a double-bend achromatic lens for pulse compression and jitter removal. The typical electron beam spot size on the sample was approximately 150~$\upmu$m measured in full-width at half maximum (FWHM), nearly five times smaller than the size of the pump pulse, ensuring a homogeneous photoexcitation condition. The temporal delay between the pump and probe pulses was adjusted by a linear translation stage. Diffracted electron beams were incident on a phosphor screen (P43) and the image was collected by an electron-multiplying charge-coupled device. In the low temperature experiments, the sample was cooled by liquid nitrogen. Further details of the UED beamline can be found in ref.~\cite{Qi2020}.\\

\noindent\textbf{Transmission electron microscopy.} The TEM experiments were conducted using a JEOL JEM-2100F microscope, where the accelerating voltage was 200~kV. The cross-section TEM sample was prepared by a focused-ion beam technique based on the standard lift-out procedure and the sample size was approximately $20~\upmu\text{m} \times 10~\upmu\text{m} \times 60~$nm. The beam spot size was 10~$\upmu$m in diameter and the current density was 0.2~pA/cm$^2$. The electron beam was incident on the sample along the direction of a straight height step, which is normally the edge of a low-index plane, and the zone axis was further confirmed by diffraction pattern simulations. Low-temperature data were collected with liquid nitrogen cooling.

\section{Acknowledgements}
We thank Weidong~Luo, Chengyang~Xu, Sheng~Meng, and Yu~He for helpful discussions. D.X. and J.Z. acknowledge support from the National Key R\&D Program of China (No.~2021YFA1400202), the National Natural Science Foundation of China (grant No.'s~11925505, 11504232 and 11721091), and the office of Science and Technology, Shanghai Municipal Government (No.'s~16DZ2260200). J.L. acknowledges support from the National Natural Science Foundation of China under grant No. 12074408. L.Y. acknowledges support from the National Natural Science Foundation of China (grant No.~11774190). W.Z. acknowledges support from the Ministry of Science and Technology of China (2016YFA0300501) and the National Natural Science Foundation of China (11974243), and additional support from a Shanghai talent program. Y.G. acknowledges support from the National Natural Science Foundation of China (grant No.~11874264). D.Q. acknowledges support from the Ministry of Science and Technology of China (grant No.~2016YFA0301003), and the National Natural Science Foundation of China (grant No.'s~12074248 and 11521404). A.Z. acknowledges support from the Miller Institute for Basic Research in Science. M.W.Z. acknowledges funding by the W.~M. Keck Foundation, funding from the UC Oﬃce of the President within the Multicampus Research Programs and Initiatives (M21PL3263), and the Hellman Fellows Fund.


\begin{thebibliography}{33}%
\makeatletter
\providecommand \@ifxundefined [1]{%
 \@ifx{#1\undefined}
}%
\providecommand \@ifnum [1]{%
 \ifnum #1\expandafter \@firstoftwo
 \else \expandafter \@secondoftwo
 \fi
}%
\providecommand \@ifx [1]{%
 \ifx #1\expandafter \@firstoftwo
 \else \expandafter \@secondoftwo
 \fi
}%
\providecommand \natexlab [1]{#1}%
\providecommand \enquote  [1]{``#1''}%
\providecommand \bibnamefont  [1]{#1}%
\providecommand \bibfnamefont [1]{#1}%
\providecommand \citenamefont [1]{#1}%
\providecommand \href@noop [0]{\@secondoftwo}%
\providecommand \href [0]{\begingroup \@sanitize@url \@href}%
\providecommand \@href[1]{\@@startlink{#1}\@@href}%
\providecommand \@@href[1]{\endgroup#1\@@endlink}%
\providecommand \@sanitize@url [0]{\catcode `\\12\catcode `\$12\catcode
  `\&12\catcode `\#12\catcode `\^12\catcode `\_12\catcode `\%12\relax}%
\providecommand \@@startlink[1]{}%
\providecommand \@@endlink[0]{}%
\providecommand \url  [0]{\begingroup\@sanitize@url \@url }%
\providecommand \@url [1]{\endgroup\@href {#1}{\urlprefix }}%
\providecommand \urlprefix  [0]{URL }%
\providecommand \Eprint [0]{\href }%
\providecommand \doibase [0]{https://doi.org/}%
\providecommand \selectlanguage [0]{\@gobble}%
\providecommand \bibinfo  [0]{\@secondoftwo}%
\providecommand \bibfield  [0]{\@secondoftwo}%
\providecommand \translation [1]{[#1]}%
\providecommand \BibitemOpen [0]{}%
\providecommand \bibitemStop [0]{}%
\providecommand \bibitemNoStop [0]{.\EOS\space}%
\providecommand \EOS [0]{\spacefactor3000\relax}%
\providecommand \BibitemShut  [1]{\csname bibitem#1\endcsname}%
\let\auto@bib@innerbib\@empty
\bibitem [{\citenamefont {van Hove}(1950)}]{VanHove1950}%
  \BibitemOpen
  \bibfield  {author} {\bibinfo {author} {\bibfnamefont {L.}~\bibnamefont {van
  Hove}},\ }\bibfield  {title} {\bibinfo {title} {{Sur L'int{\'{e}}grale de
  Configuration Pour Les Syst{\`{e}}mes De Particules {\`{A}} Une Dimension}},\
  }\href {https://doi.org/10.1016/0031-8914(50)90072-3} {\bibfield  {journal}
  {\bibinfo  {journal} {Physica}\ }\textbf {\bibinfo {volume} {16}},\ \bibinfo
  {pages} {137} (\bibinfo {year} {1950})}\BibitemShut {NoStop}%
\bibitem [{\citenamefont {Kosterlitz}\ and\ \citenamefont
  {Thouless}(1973)}]{Kosterlitz1973}%
  \BibitemOpen
  \bibfield  {author} {\bibinfo {author} {\bibfnamefont {J.~M.}\ \bibnamefont
  {Kosterlitz}}\ and\ \bibinfo {author} {\bibfnamefont {D.~J.}\ \bibnamefont
  {Thouless}},\ }\bibfield  {title} {\bibinfo {title} {{Ordering, metastability
  and phase transitions in two-dimensional systems}},\ }\href
  {https://doi.org/10.1088/0022-3719/6/7/010} {\bibfield  {journal} {\bibinfo
  {journal} {J. Phys. C: Solid State Phys.}\ }\textbf {\bibinfo {volume} {6}},\
  \bibinfo {pages} {1181} (\bibinfo {year} {1973})}\BibitemShut {NoStop}%
\bibitem [{\citenamefont {Sun}\ and\ \citenamefont {Millis}(2020)}]{Sun2020}%
  \BibitemOpen
  \bibfield  {author} {\bibinfo {author} {\bibfnamefont {Z.}~\bibnamefont
  {Sun}}\ and\ \bibinfo {author} {\bibfnamefont {A.~J.}\ \bibnamefont
  {Millis}},\ }\bibfield  {title} {\bibinfo {title} {{Transient trapping into
  metastable states in systems with competing orders}},\ }\href
  {https://doi.org/10.1103/PhysRevX.10.021028} {\bibfield  {journal} {\bibinfo
  {journal} {Phys. Rev. X}\ }\textbf {\bibinfo {volume} {10}},\ \bibinfo
  {pages} {021028} (\bibinfo {year} {2020})}\BibitemShut {NoStop}%
\bibitem [{\citenamefont {Stojchevska}\ \emph {et~al.}(2014)\citenamefont
  {Stojchevska}, \citenamefont {Vaskivskyi}, \citenamefont {Mertelj},
  \citenamefont {Kusar}, \citenamefont {Svetin}, \citenamefont {Brazovskii},\
  and\ \citenamefont {Mihailovic}}]{Stojchevska2014}%
  \BibitemOpen
  \bibfield  {author} {\bibinfo {author} {\bibfnamefont {L.}~\bibnamefont
  {Stojchevska}}, \bibinfo {author} {\bibfnamefont {I.}~\bibnamefont
  {Vaskivskyi}}, \bibinfo {author} {\bibfnamefont {T.}~\bibnamefont {Mertelj}},
  \bibinfo {author} {\bibfnamefont {P.}~\bibnamefont {Kusar}}, \bibinfo
  {author} {\bibfnamefont {D.}~\bibnamefont {Svetin}}, \bibinfo {author}
  {\bibfnamefont {S.}~\bibnamefont {Brazovskii}},\ and\ \bibinfo {author}
  {\bibfnamefont {D.}~\bibnamefont {Mihailovic}},\ }\bibfield  {title}
  {\bibinfo {title} {{Ultrafast switching to a stable hidden quantum state in
  an electronic crystal}},\ }\href {https://doi.org/10.1126/science.1241591}
  {\bibfield  {journal} {\bibinfo  {journal} {Science}\ }\textbf {\bibinfo
  {volume} {344}},\ \bibinfo {pages} {177} (\bibinfo {year}
  {2014})}\BibitemShut {NoStop}%
\bibitem [{\citenamefont {Stoica}\ \emph {et~al.}(2019)\citenamefont {Stoica},
  \citenamefont {Laanait}, \citenamefont {Dai}, \citenamefont {Hong},
  \citenamefont {Yuan}, \citenamefont {Zhang}, \citenamefont {Lei},
  \citenamefont {McCarter}, \citenamefont {Yadav}, \citenamefont {Damodaran},
  \citenamefont {Das}, \citenamefont {Stone}, \citenamefont {Karapetrova},
  \citenamefont {Walko}, \citenamefont {Zhang}, \citenamefont {Martin},
  \citenamefont {Ramesh}, \citenamefont {Chen}, \citenamefont {Wen},
  \citenamefont {Gopalan},\ and\ \citenamefont {Freeland}}]{Stoica2019}%
  \BibitemOpen
  \bibfield  {author} {\bibinfo {author} {\bibfnamefont {V.~A.}\ \bibnamefont
  {Stoica}}, \bibinfo {author} {\bibfnamefont {N.}~\bibnamefont {Laanait}},
  \bibinfo {author} {\bibfnamefont {C.}~\bibnamefont {Dai}}, \bibinfo {author}
  {\bibfnamefont {Z.}~\bibnamefont {Hong}}, \bibinfo {author} {\bibfnamefont
  {Y.}~\bibnamefont {Yuan}}, \bibinfo {author} {\bibfnamefont {Z.}~\bibnamefont
  {Zhang}}, \bibinfo {author} {\bibfnamefont {S.}~\bibnamefont {Lei}}, \bibinfo
  {author} {\bibfnamefont {M.~R.}\ \bibnamefont {McCarter}}, \bibinfo {author}
  {\bibfnamefont {A.}~\bibnamefont {Yadav}}, \bibinfo {author} {\bibfnamefont
  {A.~R.}\ \bibnamefont {Damodaran}}, \bibinfo {author} {\bibfnamefont
  {S.}~\bibnamefont {Das}}, \bibinfo {author} {\bibfnamefont {G.~A.}\
  \bibnamefont {Stone}}, \bibinfo {author} {\bibfnamefont {J.}~\bibnamefont
  {Karapetrova}}, \bibinfo {author} {\bibfnamefont {D.~A.}\ \bibnamefont
  {Walko}}, \bibinfo {author} {\bibfnamefont {X.}~\bibnamefont {Zhang}},
  \bibinfo {author} {\bibfnamefont {L.~W.}\ \bibnamefont {Martin}}, \bibinfo
  {author} {\bibfnamefont {R.}~\bibnamefont {Ramesh}}, \bibinfo {author}
  {\bibfnamefont {L.-Q.}\ \bibnamefont {Chen}}, \bibinfo {author}
  {\bibfnamefont {H.}~\bibnamefont {Wen}}, \bibinfo {author} {\bibfnamefont
  {V.}~\bibnamefont {Gopalan}},\ and\ \bibinfo {author} {\bibfnamefont {J.~W.}\
  \bibnamefont {Freeland}},\ }\bibfield  {title} {\bibinfo {title} {{Optical
  creation of a supercrystal with three-dimensional nanoscale periodicity}},\
  }\href {https://doi.org/10.1038/s41563-019-0311-x} {\bibfield  {journal}
  {\bibinfo  {journal} {Nat. Mater.}\ }\textbf {\bibinfo {volume} {18}},\
  \bibinfo {pages} {377} (\bibinfo {year} {2019})}\BibitemShut {NoStop}%
\bibitem [{\citenamefont {Karpov}\ and\ \citenamefont
  {Brazovskii}(2018)}]{Karpov2018}%
  \BibitemOpen
  \bibfield  {author} {\bibinfo {author} {\bibfnamefont {P.}~\bibnamefont
  {Karpov}}\ and\ \bibinfo {author} {\bibfnamefont {S.}~\bibnamefont
  {Brazovskii}},\ }\bibfield  {title} {\bibinfo {title} {{Modeling of networks
  and globules of charged domain walls observed in pump and pulse induced
  states}},\ }\href {https://doi.org/10.1038/s41598-018-22308-7} {\bibfield
  {journal} {\bibinfo  {journal} {Sci. Rep.}\ }\textbf {\bibinfo {volume}
  {8}},\ \bibinfo {pages} {4043} (\bibinfo {year} {2018})}\BibitemShut
  {NoStop}%
\bibitem [{\citenamefont {Chen}\ \emph {et~al.}(2015)\citenamefont {Chen},
  \citenamefont {Chan}, \citenamefont {Fang}, \citenamefont {Zhang},
  \citenamefont {Chou}, \citenamefont {Mo}, \citenamefont {Hussain},
  \citenamefont {Fedorov},\ and\ \citenamefont {Chiang}}]{Chen2015}%
  \BibitemOpen
  \bibfield  {author} {\bibinfo {author} {\bibfnamefont {P.}~\bibnamefont
  {Chen}}, \bibinfo {author} {\bibfnamefont {Y.~H.}\ \bibnamefont {Chan}},
  \bibinfo {author} {\bibfnamefont {X.~Y.}\ \bibnamefont {Fang}}, \bibinfo
  {author} {\bibfnamefont {Y.}~\bibnamefont {Zhang}}, \bibinfo {author}
  {\bibfnamefont {M.~Y.}\ \bibnamefont {Chou}}, \bibinfo {author}
  {\bibfnamefont {S.~K.}\ \bibnamefont {Mo}}, \bibinfo {author} {\bibfnamefont
  {Z.}~\bibnamefont {Hussain}}, \bibinfo {author} {\bibfnamefont {A.~V.}\
  \bibnamefont {Fedorov}},\ and\ \bibinfo {author} {\bibfnamefont {T.~C.}\
  \bibnamefont {Chiang}},\ }\bibfield  {title} {\bibinfo {title} {{Charge
  density wave transition in single-layer titanium diselenide}},\ }\href
  {https://doi.org/10.1038/ncomms9943} {\bibfield  {journal} {\bibinfo
  {journal} {Nat. Commun.}\ }\textbf {\bibinfo {volume} {6}},\ \bibinfo {pages}
  {8943} (\bibinfo {year} {2015})}\BibitemShut {NoStop}%
\bibitem [{\citenamefont {Chen}\ \emph {et~al.}(2016)\citenamefont {Chen},
  \citenamefont {Chan}, \citenamefont {Fang}, \citenamefont {Mo}, \citenamefont
  {Hussain}, \citenamefont {Fedorov}, \citenamefont {Chou},\ and\ \citenamefont
  {Chiang}}]{Chen2016}%
  \BibitemOpen
  \bibfield  {author} {\bibinfo {author} {\bibfnamefont {P.}~\bibnamefont
  {Chen}}, \bibinfo {author} {\bibfnamefont {Y.-H.}\ \bibnamefont {Chan}},
  \bibinfo {author} {\bibfnamefont {X.-Y.}\ \bibnamefont {Fang}}, \bibinfo
  {author} {\bibfnamefont {S.-K.}\ \bibnamefont {Mo}}, \bibinfo {author}
  {\bibfnamefont {Z.}~\bibnamefont {Hussain}}, \bibinfo {author} {\bibfnamefont
  {A.-V.}\ \bibnamefont {Fedorov}}, \bibinfo {author} {\bibfnamefont {M.~Y.}\
  \bibnamefont {Chou}},\ and\ \bibinfo {author} {\bibfnamefont {T.-C.}\
  \bibnamefont {Chiang}},\ }\bibfield  {title} {\bibinfo {title} {{Hidden Order
  and Dimensional Crossover of the Charge Density Waves in TiSe$_2$}},\ }\href
  {https://doi.org/10.1038/srep37910} {\bibfield  {journal} {\bibinfo
  {journal} {Sci. Rep.}\ }\textbf {\bibinfo {volume} {6}},\ \bibinfo {pages}
  {37910} (\bibinfo {year} {2016})}\BibitemShut {NoStop}%
\bibitem [{\citenamefont {Watson}\ \emph {et~al.}(2020)\citenamefont {Watson},
  \citenamefont {Rajan}, \citenamefont {Antonelli}, \citenamefont {Underwood},
  \citenamefont {Markovi{\'{c}}}, \citenamefont {Mazzola}, \citenamefont
  {Clark}, \citenamefont {Siemann}, \citenamefont {Biswas}, \citenamefont
  {Hunter}, \citenamefont {Jandura}, \citenamefont {Reichstetter},
  \citenamefont {McLaren}, \citenamefont {{Le F{\`{e}}vre}}, \citenamefont
  {Vinai},\ and\ \citenamefont {King}}]{Watson2020}%
  \BibitemOpen
  \bibfield  {author} {\bibinfo {author} {\bibfnamefont {M.~D.}\ \bibnamefont
  {Watson}}, \bibinfo {author} {\bibfnamefont {A.}~\bibnamefont {Rajan}},
  \bibinfo {author} {\bibfnamefont {T.}~\bibnamefont {Antonelli}}, \bibinfo
  {author} {\bibfnamefont {K.}~\bibnamefont {Underwood}}, \bibinfo {author}
  {\bibfnamefont {I.}~\bibnamefont {Markovi{\'{c}}}}, \bibinfo {author}
  {\bibfnamefont {F.}~\bibnamefont {Mazzola}}, \bibinfo {author} {\bibfnamefont
  {O.~J.}\ \bibnamefont {Clark}}, \bibinfo {author} {\bibfnamefont {G.-R.}\
  \bibnamefont {Siemann}}, \bibinfo {author} {\bibfnamefont {D.}~\bibnamefont
  {Biswas}}, \bibinfo {author} {\bibfnamefont {A.}~\bibnamefont {Hunter}},
  \bibinfo {author} {\bibfnamefont {S.}~\bibnamefont {Jandura}}, \bibinfo
  {author} {\bibfnamefont {J.}~\bibnamefont {Reichstetter}}, \bibinfo {author}
  {\bibfnamefont {M.}~\bibnamefont {McLaren}}, \bibinfo {author} {\bibfnamefont
  {P.}~\bibnamefont {{Le F{\`{e}}vre}}}, \bibinfo {author} {\bibfnamefont
  {G.}~\bibnamefont {Vinai}},\ and\ \bibinfo {author} {\bibfnamefont
  {P.~D.~C.}\ \bibnamefont {King}},\ }\bibfield  {title} {\bibinfo {title}
  {{Strong-coupling charge density wave in monolayer TiSe$_2$}},\ }\href
  {https://doi.org/10.1088/2053-1583/abafec} {\bibfield  {journal} {\bibinfo
  {journal} {2D Mater.}\ }\textbf {\bibinfo {volume} {8}},\ \bibinfo {pages}
  {015004} (\bibinfo {year} {2020})}\BibitemShut {NoStop}%
\bibitem [{\citenamefont {Kogar}\ \emph {et~al.}(2017)\citenamefont {Kogar},
  \citenamefont {Rak}, \citenamefont {Vig}, \citenamefont {Husain},
  \citenamefont {Flicker}, \citenamefont {Joe}, \citenamefont {Venema},
  \citenamefont {MacDougall}, \citenamefont {Chiang}, \citenamefont {Fradkin},
  \citenamefont {van Wezel},\ and\ \citenamefont {Abbamonte}}]{Kogar2017}%
  \BibitemOpen
  \bibfield  {author} {\bibinfo {author} {\bibfnamefont {A.}~\bibnamefont
  {Kogar}}, \bibinfo {author} {\bibfnamefont {M.~S.}\ \bibnamefont {Rak}},
  \bibinfo {author} {\bibfnamefont {S.}~\bibnamefont {Vig}}, \bibinfo {author}
  {\bibfnamefont {A.~A.}\ \bibnamefont {Husain}}, \bibinfo {author}
  {\bibfnamefont {F.}~\bibnamefont {Flicker}}, \bibinfo {author} {\bibfnamefont
  {Y.~I.}\ \bibnamefont {Joe}}, \bibinfo {author} {\bibfnamefont
  {L.}~\bibnamefont {Venema}}, \bibinfo {author} {\bibfnamefont {G.~J.}\
  \bibnamefont {MacDougall}}, \bibinfo {author} {\bibfnamefont {T.~C.}\
  \bibnamefont {Chiang}}, \bibinfo {author} {\bibfnamefont {E.}~\bibnamefont
  {Fradkin}}, \bibinfo {author} {\bibfnamefont {J.}~\bibnamefont {van Wezel}},\
  and\ \bibinfo {author} {\bibfnamefont {P.}~\bibnamefont {Abbamonte}},\
  }\bibfield  {title} {\bibinfo {title} {{Signatures of exciton condensation in
  a transition metal dichalcogenide}},\ }\href
  {https://doi.org/10.1126/science.aam6432} {\bibfield  {journal} {\bibinfo
  {journal} {Science}\ }\textbf {\bibinfo {volume} {358}},\ \bibinfo {pages}
  {1314} (\bibinfo {year} {2017})}\BibitemShut {NoStop}%
\bibitem [{\citenamefont {Hughes}(1977)}]{Hughes1977}%
  \BibitemOpen
  \bibfield  {author} {\bibinfo {author} {\bibfnamefont {H.~P.}\ \bibnamefont
  {Hughes}},\ }\bibfield  {title} {\bibinfo {title} {{Structural distortion in
  TiSe$_2$ and related materials -- a possible Jahn-Teller effect?}},\ }\href
  {https://doi.org/10.1088/0022-3719/10/11/009} {\bibfield  {journal} {\bibinfo
   {journal} {J. Phys. C: Solid State Phys.}\ }\textbf {\bibinfo {volume}
  {10}},\ \bibinfo {pages} {L319} (\bibinfo {year} {1977})}\BibitemShut
  {NoStop}%
\bibitem [{\citenamefont {Suzuki}\ \emph {et~al.}(1985)\citenamefont {Suzuki},
  \citenamefont {Yamamoto},\ and\ \citenamefont {Motizuki}}]{Suzuki1985}%
  \BibitemOpen
  \bibfield  {author} {\bibinfo {author} {\bibfnamefont {N.}~\bibnamefont
  {Suzuki}}, \bibinfo {author} {\bibfnamefont {A.}~\bibnamefont {Yamamoto}},\
  and\ \bibinfo {author} {\bibfnamefont {K.}~\bibnamefont {Motizuki}},\
  }\bibfield  {title} {\bibinfo {title} {{Microscopic theory of the CDW state
  of 1$T$-TiSe$_2$}},\ }\href {https://doi.org/10.1143/JPSJ.54.4668} {\bibfield
   {journal} {\bibinfo  {journal} {J. Phys. Soc. Japan}\ }\textbf {\bibinfo
  {volume} {54}},\ \bibinfo {pages} {4668} (\bibinfo {year}
  {1985})}\BibitemShut {NoStop}%
\bibitem [{\citenamefont {Kidd}\ \emph {et~al.}(2002)\citenamefont {Kidd},
  \citenamefont {Miller}, \citenamefont {Chou},\ and\ \citenamefont
  {Chiang}}]{Kidd2002}%
  \BibitemOpen
  \bibfield  {author} {\bibinfo {author} {\bibfnamefont {T.~E.}\ \bibnamefont
  {Kidd}}, \bibinfo {author} {\bibfnamefont {T.}~\bibnamefont {Miller}},
  \bibinfo {author} {\bibfnamefont {M.~Y.}\ \bibnamefont {Chou}},\ and\
  \bibinfo {author} {\bibfnamefont {T.-C.}\ \bibnamefont {Chiang}},\ }\bibfield
   {title} {\bibinfo {title} {{Electron-hole coupling and the charge density
  wave transition in TiSe$_2$}},\ }\href
  {https://doi.org/10.1103/PhysRevLett.88.226402} {\bibfield  {journal}
  {\bibinfo  {journal} {Phys. Rev. Lett.}\ }\textbf {\bibinfo {volume} {88}},\
  \bibinfo {pages} {226402} (\bibinfo {year} {2002})}\BibitemShut {NoStop}%
\bibitem [{\citenamefont {Cercellier}\ \emph {et~al.}(2007)\citenamefont
  {Cercellier}, \citenamefont {Monney}, \citenamefont {Clerc}, \citenamefont
  {Battaglia}, \citenamefont {Despont}, \citenamefont {Garnier}, \citenamefont
  {Beck}, \citenamefont {Aebi}, \citenamefont {Patthey}, \citenamefont
  {Berger},\ and\ \citenamefont {Forr{\'{o}}}}]{Cercellier2007}%
  \BibitemOpen
  \bibfield  {author} {\bibinfo {author} {\bibfnamefont {H.}~\bibnamefont
  {Cercellier}}, \bibinfo {author} {\bibfnamefont {C.}~\bibnamefont {Monney}},
  \bibinfo {author} {\bibfnamefont {F.}~\bibnamefont {Clerc}}, \bibinfo
  {author} {\bibfnamefont {C.}~\bibnamefont {Battaglia}}, \bibinfo {author}
  {\bibfnamefont {L.}~\bibnamefont {Despont}}, \bibinfo {author} {\bibfnamefont
  {M.~G.}\ \bibnamefont {Garnier}}, \bibinfo {author} {\bibfnamefont
  {H.}~\bibnamefont {Beck}}, \bibinfo {author} {\bibfnamefont {P.}~\bibnamefont
  {Aebi}}, \bibinfo {author} {\bibfnamefont {L.}~\bibnamefont {Patthey}},
  \bibinfo {author} {\bibfnamefont {H.}~\bibnamefont {Berger}},\ and\ \bibinfo
  {author} {\bibfnamefont {L.}~\bibnamefont {Forr{\'{o}}}},\ }\bibfield
  {title} {\bibinfo {title} {{Evidence for an excitonic insulator phase in
  1$T$-TiSe$_2$}},\ }\href {https://doi.org/10.1103/PhysRevLett.99.146403}
  {\bibfield  {journal} {\bibinfo  {journal} {Phys. Rev. Lett.}\ }\textbf
  {\bibinfo {volume} {99}},\ \bibinfo {pages} {146403} (\bibinfo {year}
  {2007})}\BibitemShut {NoStop}%
\bibitem [{\citenamefont {Calandra}\ and\ \citenamefont
  {Mauri}(2011)}]{Calandra2011}%
  \BibitemOpen
  \bibfield  {author} {\bibinfo {author} {\bibfnamefont {M.}~\bibnamefont
  {Calandra}}\ and\ \bibinfo {author} {\bibfnamefont {F.}~\bibnamefont
  {Mauri}},\ }\bibfield  {title} {\bibinfo {title} {{Charge-density wave and
  superconducting dome in TiSe$_2$ from electron-phonon interaction}},\ }\href
  {https://doi.org/10.1103/PhysRevLett.106.196406} {\bibfield  {journal}
  {\bibinfo  {journal} {Phys. Rev. Lett.}\ }\textbf {\bibinfo {volume} {106}},\
  \bibinfo {pages} {196406} (\bibinfo {year} {2011})}\BibitemShut {NoStop}%
\bibitem [{\citenamefont {Porer}\ \emph {et~al.}(2014)\citenamefont {Porer},
  \citenamefont {Leierseder}, \citenamefont {M{\'{e}}nard}, \citenamefont
  {Dachraoui}, \citenamefont {Mouchliadis}, \citenamefont {Perakis},
  \citenamefont {Heinzmann}, \citenamefont {Demsar}, \citenamefont
  {Rossnagel},\ and\ \citenamefont {Huber}}]{Porer2014}%
  \BibitemOpen
  \bibfield  {author} {\bibinfo {author} {\bibfnamefont {M.}~\bibnamefont
  {Porer}}, \bibinfo {author} {\bibfnamefont {U.}~\bibnamefont {Leierseder}},
  \bibinfo {author} {\bibfnamefont {J.-M.}\ \bibnamefont {M{\'{e}}nard}},
  \bibinfo {author} {\bibfnamefont {H.}~\bibnamefont {Dachraoui}}, \bibinfo
  {author} {\bibfnamefont {L.}~\bibnamefont {Mouchliadis}}, \bibinfo {author}
  {\bibfnamefont {I.~E.}\ \bibnamefont {Perakis}}, \bibinfo {author}
  {\bibfnamefont {U.}~\bibnamefont {Heinzmann}}, \bibinfo {author}
  {\bibfnamefont {J.}~\bibnamefont {Demsar}}, \bibinfo {author} {\bibfnamefont
  {K.}~\bibnamefont {Rossnagel}},\ and\ \bibinfo {author} {\bibfnamefont
  {R.}~\bibnamefont {Huber}},\ }\bibfield  {title} {\bibinfo {title}
  {{Non-thermal separation of electronic and structural orders in a persisting
  charge density wave}},\ }\href {https://doi.org/10.1038/nmat4042} {\bibfield
  {journal} {\bibinfo  {journal} {Nat. Mater.}\ }\textbf {\bibinfo {volume}
  {13}},\ \bibinfo {pages} {857} (\bibinfo {year} {2014})}\BibitemShut
  {NoStop}%
\bibitem [{\citenamefont {Burian}\ \emph {et~al.}(2021)\citenamefont {Burian},
  \citenamefont {Porer}, \citenamefont {Mardegan}, \citenamefont {Esposito},
  \citenamefont {Parchenko}, \citenamefont {Burganov}, \citenamefont {Gurung},
  \citenamefont {Ramakrishnan}, \citenamefont {Scagnoli}, \citenamefont {Ueda},
  \citenamefont {Francoual}, \citenamefont {Fabrizi}, \citenamefont {Tanaka},
  \citenamefont {Togashi}, \citenamefont {Kubota}, \citenamefont {Yabashi},
  \citenamefont {Rossnagel}, \citenamefont {Johnson},\ and\ \citenamefont
  {Staub}}]{Burian2021}%
  \BibitemOpen
  \bibfield  {author} {\bibinfo {author} {\bibfnamefont {M.}~\bibnamefont
  {Burian}}, \bibinfo {author} {\bibfnamefont {M.}~\bibnamefont {Porer}},
  \bibinfo {author} {\bibfnamefont {J.~R.~L.}\ \bibnamefont {Mardegan}},
  \bibinfo {author} {\bibfnamefont {V.}~\bibnamefont {Esposito}}, \bibinfo
  {author} {\bibfnamefont {S.}~\bibnamefont {Parchenko}}, \bibinfo {author}
  {\bibfnamefont {B.}~\bibnamefont {Burganov}}, \bibinfo {author}
  {\bibfnamefont {N.}~\bibnamefont {Gurung}}, \bibinfo {author} {\bibfnamefont
  {M.}~\bibnamefont {Ramakrishnan}}, \bibinfo {author} {\bibfnamefont
  {V.}~\bibnamefont {Scagnoli}}, \bibinfo {author} {\bibfnamefont
  {H.}~\bibnamefont {Ueda}}, \bibinfo {author} {\bibfnamefont {S.}~\bibnamefont
  {Francoual}}, \bibinfo {author} {\bibfnamefont {F.}~\bibnamefont {Fabrizi}},
  \bibinfo {author} {\bibfnamefont {Y.}~\bibnamefont {Tanaka}}, \bibinfo
  {author} {\bibfnamefont {T.}~\bibnamefont {Togashi}}, \bibinfo {author}
  {\bibfnamefont {Y.}~\bibnamefont {Kubota}}, \bibinfo {author} {\bibfnamefont
  {M.}~\bibnamefont {Yabashi}}, \bibinfo {author} {\bibfnamefont
  {K.}~\bibnamefont {Rossnagel}}, \bibinfo {author} {\bibfnamefont {S.~L.}\
  \bibnamefont {Johnson}},\ and\ \bibinfo {author} {\bibfnamefont
  {U.}~\bibnamefont {Staub}},\ }\bibfield  {title} {\bibinfo {title}
  {{Structural involvement in the melting of the charge density wave in
  1$T$-TiSe$_2$}},\ }\href {https://doi.org/10.1103/PhysRevResearch.3.013128}
  {\bibfield  {journal} {\bibinfo  {journal} {Phys. Rev. Research}\ }\textbf
  {\bibinfo {volume} {3}},\ \bibinfo {pages} {013128} (\bibinfo {year}
  {2021})}\BibitemShut {NoStop}%
\bibitem [{\citenamefont {Bie}\ \emph {et~al.}(2021)\citenamefont {Bie},
  \citenamefont {Zong}, \citenamefont {Wang}, \citenamefont {Jarillo-Herrero},\
  and\ \citenamefont {Gedik}}]{Bie2021}%
  \BibitemOpen
  \bibfield  {author} {\bibinfo {author} {\bibfnamefont {Y.-Q.}\ \bibnamefont
  {Bie}}, \bibinfo {author} {\bibfnamefont {A.}~\bibnamefont {Zong}}, \bibinfo
  {author} {\bibfnamefont {X.}~\bibnamefont {Wang}}, \bibinfo {author}
  {\bibfnamefont {P.}~\bibnamefont {Jarillo-Herrero}},\ and\ \bibinfo {author}
  {\bibfnamefont {N.}~\bibnamefont {Gedik}},\ }\bibfield  {title} {\bibinfo
  {title} {{A versatile sample fabrication method for ultrafast electron
  diffraction}},\ }\href {https://doi.org/10.1016/j.ultramic.2021.113389}
  {\bibfield  {journal} {\bibinfo  {journal} {Ultramicroscopy}\ }\textbf
  {\bibinfo {volume} {230}},\ \bibinfo {pages} {113389} (\bibinfo {year}
  {2021})}\BibitemShut {NoStop}%
\bibitem [{\citenamefont {Huber}\ \emph {et~al.}(2014)\citenamefont {Huber},
  \citenamefont {Mariager}, \citenamefont {Ferrer}, \citenamefont
  {Sch{\"{a}}fer}, \citenamefont {Johnson}, \citenamefont {Gr{\"{u}}bel},
  \citenamefont {L{\"{u}}bcke}, \citenamefont {Huber}, \citenamefont {Kubacka},
  \citenamefont {Dornes}, \citenamefont {Laulhe}, \citenamefont {Ravy},
  \citenamefont {Ingold}, \citenamefont {Beaud}, \citenamefont {Demsar},\ and\
  \citenamefont {Johnson}}]{Huber2014}%
  \BibitemOpen
  \bibfield  {author} {\bibinfo {author} {\bibfnamefont {T.}~\bibnamefont
  {Huber}}, \bibinfo {author} {\bibfnamefont {S.~O.}\ \bibnamefont {Mariager}},
  \bibinfo {author} {\bibfnamefont {A.}~\bibnamefont {Ferrer}}, \bibinfo
  {author} {\bibfnamefont {H.}~\bibnamefont {Sch{\"{a}}fer}}, \bibinfo {author}
  {\bibfnamefont {J.~A.}\ \bibnamefont {Johnson}}, \bibinfo {author}
  {\bibfnamefont {S.}~\bibnamefont {Gr{\"{u}}bel}}, \bibinfo {author}
  {\bibfnamefont {A.}~\bibnamefont {L{\"{u}}bcke}}, \bibinfo {author}
  {\bibfnamefont {L.}~\bibnamefont {Huber}}, \bibinfo {author} {\bibfnamefont
  {T.}~\bibnamefont {Kubacka}}, \bibinfo {author} {\bibfnamefont
  {C.}~\bibnamefont {Dornes}}, \bibinfo {author} {\bibfnamefont
  {C.}~\bibnamefont {Laulhe}}, \bibinfo {author} {\bibfnamefont
  {S.}~\bibnamefont {Ravy}}, \bibinfo {author} {\bibfnamefont {G.}~\bibnamefont
  {Ingold}}, \bibinfo {author} {\bibfnamefont {P.}~\bibnamefont {Beaud}},
  \bibinfo {author} {\bibfnamefont {J.}~\bibnamefont {Demsar}},\ and\ \bibinfo
  {author} {\bibfnamefont {S.~L.}\ \bibnamefont {Johnson}},\ }\bibfield
  {title} {\bibinfo {title} {{Coherent structural dynamics of a prototypical
  charge-density-wave-to-metal transition}},\ }\href
  {https://doi.org/10.1103/PhysRevLett.113.026401} {\bibfield  {journal}
  {\bibinfo  {journal} {Phys. Rev. Lett.}\ }\textbf {\bibinfo {volume} {113}},\
  \bibinfo {pages} {026401} (\bibinfo {year} {2014})}\BibitemShut {NoStop}%
\bibitem [{\citenamefont {Moore}\ \emph {et~al.}(2016)\citenamefont {Moore},
  \citenamefont {Lee}, \citenamefont {Kirchman}, \citenamefont {Chuang},
  \citenamefont {Kemper}, \citenamefont {Trigo}, \citenamefont {Patthey},
  \citenamefont {Lu}, \citenamefont {Krupin}, \citenamefont {Yi}, \citenamefont
  {Reis}, \citenamefont {Doering}, \citenamefont {Denes}, \citenamefont
  {Schlotter}, \citenamefont {Turner}, \citenamefont {Hays}, \citenamefont
  {Hering}, \citenamefont {Benson}, \citenamefont {Chu}, \citenamefont
  {Devereaux}, \citenamefont {Fisher}, \citenamefont {Hussain},\ and\
  \citenamefont {Shen}}]{Moore2016}%
  \BibitemOpen
  \bibfield  {author} {\bibinfo {author} {\bibfnamefont {R.~G.}\ \bibnamefont
  {Moore}}, \bibinfo {author} {\bibfnamefont {W.~S.}\ \bibnamefont {Lee}},
  \bibinfo {author} {\bibfnamefont {P.~S.}\ \bibnamefont {Kirchman}}, \bibinfo
  {author} {\bibfnamefont {Y.~D.}\ \bibnamefont {Chuang}}, \bibinfo {author}
  {\bibfnamefont {A.~F.}\ \bibnamefont {Kemper}}, \bibinfo {author}
  {\bibfnamefont {M.}~\bibnamefont {Trigo}}, \bibinfo {author} {\bibfnamefont
  {L.}~\bibnamefont {Patthey}}, \bibinfo {author} {\bibfnamefont {D.~H.}\
  \bibnamefont {Lu}}, \bibinfo {author} {\bibfnamefont {O.}~\bibnamefont
  {Krupin}}, \bibinfo {author} {\bibfnamefont {M.}~\bibnamefont {Yi}}, \bibinfo
  {author} {\bibfnamefont {D.~A.}\ \bibnamefont {Reis}}, \bibinfo {author}
  {\bibfnamefont {D.}~\bibnamefont {Doering}}, \bibinfo {author} {\bibfnamefont
  {P.}~\bibnamefont {Denes}}, \bibinfo {author} {\bibfnamefont {W.~F.}\
  \bibnamefont {Schlotter}}, \bibinfo {author} {\bibfnamefont {J.~J.}\
  \bibnamefont {Turner}}, \bibinfo {author} {\bibfnamefont {G.}~\bibnamefont
  {Hays}}, \bibinfo {author} {\bibfnamefont {P.}~\bibnamefont {Hering}},
  \bibinfo {author} {\bibfnamefont {T.}~\bibnamefont {Benson}}, \bibinfo
  {author} {\bibfnamefont {J.-H.}\ \bibnamefont {Chu}}, \bibinfo {author}
  {\bibfnamefont {T.~P.}\ \bibnamefont {Devereaux}}, \bibinfo {author}
  {\bibfnamefont {I.~R.}\ \bibnamefont {Fisher}}, \bibinfo {author}
  {\bibfnamefont {Z.}~\bibnamefont {Hussain}},\ and\ \bibinfo {author}
  {\bibfnamefont {Z.-X.}\ \bibnamefont {Shen}},\ }\bibfield  {title} {\bibinfo
  {title} {{Ultrafast resonant soft X-ray diffraction dynamics of the charge
  density wave in TbTe$_3$}},\ }\href
  {https://doi.org/10.1103/PhysRevB.93.024304} {\bibfield  {journal} {\bibinfo
  {journal} {Phys. Rev. B}\ }\textbf {\bibinfo {volume} {93}},\ \bibinfo
  {pages} {024304} (\bibinfo {year} {2016})}\BibitemShut {NoStop}%
\bibitem [{\citenamefont {Wall}\ \emph {et~al.}(2018)\citenamefont {Wall},
  \citenamefont {Yang}, \citenamefont {Vidas}, \citenamefont {Chollet},
  \citenamefont {Glownia}, \citenamefont {Kozina}, \citenamefont {Katayama},
  \citenamefont {Henighan}, \citenamefont {Jiang}, \citenamefont {Miller},
  \citenamefont {Reis}, \citenamefont {Boatner}, \citenamefont {Delaire},\ and\
  \citenamefont {Trigo}}]{Wall2018}%
  \BibitemOpen
  \bibfield  {author} {\bibinfo {author} {\bibfnamefont {S.}~\bibnamefont
  {Wall}}, \bibinfo {author} {\bibfnamefont {S.}~\bibnamefont {Yang}}, \bibinfo
  {author} {\bibfnamefont {L.}~\bibnamefont {Vidas}}, \bibinfo {author}
  {\bibfnamefont {M.}~\bibnamefont {Chollet}}, \bibinfo {author} {\bibfnamefont
  {J.~M.}\ \bibnamefont {Glownia}}, \bibinfo {author} {\bibfnamefont
  {M.}~\bibnamefont {Kozina}}, \bibinfo {author} {\bibfnamefont
  {T.}~\bibnamefont {Katayama}}, \bibinfo {author} {\bibfnamefont
  {T.}~\bibnamefont {Henighan}}, \bibinfo {author} {\bibfnamefont
  {M.}~\bibnamefont {Jiang}}, \bibinfo {author} {\bibfnamefont {T.~A.}\
  \bibnamefont {Miller}}, \bibinfo {author} {\bibfnamefont {D.~A.}\
  \bibnamefont {Reis}}, \bibinfo {author} {\bibfnamefont {L.~A.}\ \bibnamefont
  {Boatner}}, \bibinfo {author} {\bibfnamefont {O.}~\bibnamefont {Delaire}},\
  and\ \bibinfo {author} {\bibfnamefont {M.}~\bibnamefont {Trigo}},\ }\bibfield
   {title} {\bibinfo {title} {{Ultrafast disordering of vanadium dimers in
  photoexcited VO$_2$}},\ }\href {https://doi.org/10.1126/science.aau3873}
  {\bibfield  {journal} {\bibinfo  {journal} {Science}\ }\textbf {\bibinfo
  {volume} {362}},\ \bibinfo {pages} {572} (\bibinfo {year}
  {2018})}\BibitemShut {NoStop}%
\bibitem [{\citenamefont {Zong}\ \emph {et~al.}(2021)\citenamefont {Zong},
  \citenamefont {Dolgirev}, \citenamefont {Kogar}, \citenamefont {Su},
  \citenamefont {Shen}, \citenamefont {Straquadine}, \citenamefont {Wang},
  \citenamefont {Luo}, \citenamefont {Kozina}, \citenamefont {Reid},
  \citenamefont {Li}, \citenamefont {Yang}, \citenamefont {Weathersby},
  \citenamefont {Park}, \citenamefont {Sie}, \citenamefont {Jarillo-Herrero},
  \citenamefont {Fisher}, \citenamefont {Wang}, \citenamefont {Demler},\ and\
  \citenamefont {Gedik}}]{Zong2021}%
  \BibitemOpen
  \bibfield  {author} {\bibinfo {author} {\bibfnamefont {A.}~\bibnamefont
  {Zong}}, \bibinfo {author} {\bibfnamefont {P.~E.}\ \bibnamefont {Dolgirev}},
  \bibinfo {author} {\bibfnamefont {A.}~\bibnamefont {Kogar}}, \bibinfo
  {author} {\bibfnamefont {Y.}~\bibnamefont {Su}}, \bibinfo {author}
  {\bibfnamefont {X.}~\bibnamefont {Shen}}, \bibinfo {author} {\bibfnamefont
  {J.~A.~W.}\ \bibnamefont {Straquadine}}, \bibinfo {author} {\bibfnamefont
  {X.}~\bibnamefont {Wang}}, \bibinfo {author} {\bibfnamefont {D.}~\bibnamefont
  {Luo}}, \bibinfo {author} {\bibfnamefont {M.~E.}\ \bibnamefont {Kozina}},
  \bibinfo {author} {\bibfnamefont {A.~H.}\ \bibnamefont {Reid}}, \bibinfo
  {author} {\bibfnamefont {R.}~\bibnamefont {Li}}, \bibinfo {author}
  {\bibfnamefont {J.}~\bibnamefont {Yang}}, \bibinfo {author} {\bibfnamefont
  {S.~P.}\ \bibnamefont {Weathersby}}, \bibinfo {author} {\bibfnamefont
  {S.}~\bibnamefont {Park}}, \bibinfo {author} {\bibfnamefont {E.~J.}\
  \bibnamefont {Sie}}, \bibinfo {author} {\bibfnamefont {P.}~\bibnamefont
  {Jarillo-Herrero}}, \bibinfo {author} {\bibfnamefont {I.~R.}\ \bibnamefont
  {Fisher}}, \bibinfo {author} {\bibfnamefont {X.}~\bibnamefont {Wang}},
  \bibinfo {author} {\bibfnamefont {E.}~\bibnamefont {Demler}},\ and\ \bibinfo
  {author} {\bibfnamefont {N.}~\bibnamefont {Gedik}},\ }\bibfield  {title}
  {\bibinfo {title} {{Role of Equilibrium Fluctuations in Light-Induced
  Order}},\ }\href {https://doi.org/10.1103/PhysRevLett.127.227401} {\bibfield
  {journal} {\bibinfo  {journal} {Phys. Rev. Lett.}\ }\textbf {\bibinfo
  {volume} {127}},\ \bibinfo {pages} {227401} (\bibinfo {year}
  {2021})}\BibitemShut {NoStop}%
\bibitem [{\citenamefont {Ravnik}\ \emph {et~al.}(2018)\citenamefont {Ravnik},
  \citenamefont {Vaskivskyi}, \citenamefont {Mertelj},\ and\ \citenamefont
  {Mihailovic}}]{Ravnik2018}%
  \BibitemOpen
  \bibfield  {author} {\bibinfo {author} {\bibfnamefont {J.}~\bibnamefont
  {Ravnik}}, \bibinfo {author} {\bibfnamefont {I.}~\bibnamefont {Vaskivskyi}},
  \bibinfo {author} {\bibfnamefont {T.}~\bibnamefont {Mertelj}},\ and\ \bibinfo
  {author} {\bibfnamefont {D.}~\bibnamefont {Mihailovic}},\ }\bibfield  {title}
  {\bibinfo {title} {{Real-time observation of the coherent transition to a
  metastable emergent state in 1$T$-TaS$_2$}},\ }\href
  {https://doi.org/10.1103/PhysRevB.97.075304} {\bibfield  {journal} {\bibinfo
  {journal} {Phys. Rev. B}\ }\textbf {\bibinfo {volume} {97}},\ \bibinfo
  {pages} {075304} (\bibinfo {year} {2018})}\BibitemShut {NoStop}%
\bibitem [{\citenamefont {Vogelgesang}\ \emph {et~al.}(2018)\citenamefont
  {Vogelgesang}, \citenamefont {Storeck}, \citenamefont {Horstmann},
  \citenamefont {Diekmann}, \citenamefont {Sivis}, \citenamefont {Schramm},
  \citenamefont {Rossnagel}, \citenamefont {Sch{\"a}fer},\ and\ \citenamefont
  {Ropers}}]{Vogelgesang2018}%
  \BibitemOpen
  \bibfield  {author} {\bibinfo {author} {\bibfnamefont {S.}~\bibnamefont
  {Vogelgesang}}, \bibinfo {author} {\bibfnamefont {G.}~\bibnamefont
  {Storeck}}, \bibinfo {author} {\bibfnamefont {J.~G.}\ \bibnamefont
  {Horstmann}}, \bibinfo {author} {\bibfnamefont {T.}~\bibnamefont {Diekmann}},
  \bibinfo {author} {\bibfnamefont {M.}~\bibnamefont {Sivis}}, \bibinfo
  {author} {\bibfnamefont {S.}~\bibnamefont {Schramm}}, \bibinfo {author}
  {\bibfnamefont {K.}~\bibnamefont {Rossnagel}}, \bibinfo {author}
  {\bibfnamefont {S.}~\bibnamefont {Sch{\"a}fer}},\ and\ \bibinfo {author}
  {\bibfnamefont {C.}~\bibnamefont {Ropers}},\ }\bibfield  {title} {\bibinfo
  {title} {{Phase ordering of charge density waves traced by ultrafast
  low-energy electron diffraction}},\ }\href
  {https://doi.org/10.1038/nphys4309} {\bibfield  {journal} {\bibinfo
  {journal} {Nat. Phys.}\ }\textbf {\bibinfo {volume} {14}},\ \bibinfo {pages}
  {184} (\bibinfo {year} {2018})}\BibitemShut {NoStop}%
\bibitem [{\citenamefont {Zong}\ \emph {et~al.}(2019)\citenamefont {Zong},
  \citenamefont {Kogar}, \citenamefont {Bie}, \citenamefont {Rohwer},
  \citenamefont {Lee}, \citenamefont {Baldini}, \citenamefont {Erge{\c{c}}en},
  \citenamefont {Yilmaz}, \citenamefont {Freelon}, \citenamefont {Sie},
  \citenamefont {Zhou}, \citenamefont {Straquadine}, \citenamefont {Walmsley},
  \citenamefont {Dolgirev}, \citenamefont {Rozhkov}, \citenamefont {Fisher},
  \citenamefont {Jarillo-Herrero}, \citenamefont {Fine},\ and\ \citenamefont
  {Gedik}}]{Zong2019}%
  \BibitemOpen
  \bibfield  {author} {\bibinfo {author} {\bibfnamefont {A.}~\bibnamefont
  {Zong}}, \bibinfo {author} {\bibfnamefont {A.}~\bibnamefont {Kogar}},
  \bibinfo {author} {\bibfnamefont {Y.-Q.}\ \bibnamefont {Bie}}, \bibinfo
  {author} {\bibfnamefont {T.}~\bibnamefont {Rohwer}}, \bibinfo {author}
  {\bibfnamefont {C.}~\bibnamefont {Lee}}, \bibinfo {author} {\bibfnamefont
  {E.}~\bibnamefont {Baldini}}, \bibinfo {author} {\bibfnamefont
  {E.}~\bibnamefont {Erge{\c{c}}en}}, \bibinfo {author} {\bibfnamefont {M.~B.}\
  \bibnamefont {Yilmaz}}, \bibinfo {author} {\bibfnamefont {B.}~\bibnamefont
  {Freelon}}, \bibinfo {author} {\bibfnamefont {E.~J.}\ \bibnamefont {Sie}},
  \bibinfo {author} {\bibfnamefont {H.}~\bibnamefont {Zhou}}, \bibinfo {author}
  {\bibfnamefont {J.}~\bibnamefont {Straquadine}}, \bibinfo {author}
  {\bibfnamefont {P.}~\bibnamefont {Walmsley}}, \bibinfo {author}
  {\bibfnamefont {P.~E.}\ \bibnamefont {Dolgirev}}, \bibinfo {author}
  {\bibfnamefont {A.~V.}\ \bibnamefont {Rozhkov}}, \bibinfo {author}
  {\bibfnamefont {I.~R.}\ \bibnamefont {Fisher}}, \bibinfo {author}
  {\bibfnamefont {P.}~\bibnamefont {Jarillo-Herrero}}, \bibinfo {author}
  {\bibfnamefont {B.~V.}\ \bibnamefont {Fine}},\ and\ \bibinfo {author}
  {\bibfnamefont {N.}~\bibnamefont {Gedik}},\ }\bibfield  {title} {\bibinfo
  {title} {{Evidence for topological defects in a photoinduced phase
  transition}},\ }\href {https://doi.org/10.1038/s41567-018-0311-9} {\bibfield
  {journal} {\bibinfo  {journal} {Nat. Phys.}\ }\textbf {\bibinfo {volume}
  {15}},\ \bibinfo {pages} {27} (\bibinfo {year} {2019})}\BibitemShut {NoStop}%
\bibitem [{\citenamefont {Otto}\ \emph {et~al.}(2021)\citenamefont {Otto},
  \citenamefont {P{\"{o}}hls}, \citenamefont {{Ren{\'{e}} de Cotret}},
  \citenamefont {Stern}, \citenamefont {Sutton},\ and\ \citenamefont
  {Siwick}}]{Otto2021}%
  \BibitemOpen
  \bibfield  {author} {\bibinfo {author} {\bibfnamefont {M.~R.}\ \bibnamefont
  {Otto}}, \bibinfo {author} {\bibfnamefont {J.-H.}\ \bibnamefont
  {P{\"{o}}hls}}, \bibinfo {author} {\bibfnamefont {L.~P.}\ \bibnamefont
  {{Ren{\'{e}} de Cotret}}}, \bibinfo {author} {\bibfnamefont {M.~J.}\
  \bibnamefont {Stern}}, \bibinfo {author} {\bibfnamefont {M.}~\bibnamefont
  {Sutton}},\ and\ \bibinfo {author} {\bibfnamefont {B.~J.}\ \bibnamefont
  {Siwick}},\ }\bibfield  {title} {\bibinfo {title} {{Mechanisms of
  electron-phonon coupling unraveled in momentum and time: The case of soft
  phonons in TiSe$_2$}},\ }\href {https://doi.org/10.1126/sciadv.abf2810}
  {\bibfield  {journal} {\bibinfo  {journal} {Sci. Adv.}\ }\textbf {\bibinfo
  {volume} {7}},\ \bibinfo {pages} {eabf2810} (\bibinfo {year}
  {2021})}\BibitemShut {NoStop}%
\bibitem [{\citenamefont {Holt}\ \emph {et~al.}(2001)\citenamefont {Holt},
  \citenamefont {Zschack}, \citenamefont {Hong}, \citenamefont {Chou},\ and\
  \citenamefont {Chiang}}]{Holt2001}%
  \BibitemOpen
  \bibfield  {author} {\bibinfo {author} {\bibfnamefont {M.}~\bibnamefont
  {Holt}}, \bibinfo {author} {\bibfnamefont {P.}~\bibnamefont {Zschack}},
  \bibinfo {author} {\bibfnamefont {H.}~\bibnamefont {Hong}}, \bibinfo {author}
  {\bibfnamefont {M.~Y.}\ \bibnamefont {Chou}},\ and\ \bibinfo {author}
  {\bibfnamefont {T.-C.}\ \bibnamefont {Chiang}},\ }\bibfield  {title}
  {\bibinfo {title} {{X-ray studies of phonon softening in TiSe$_2$}},\ }\href
  {https://doi.org/10.1103/PhysRevLett.86.3799} {\bibfield  {journal} {\bibinfo
   {journal} {Phys. Rev. Lett.}\ }\textbf {\bibinfo {volume} {86}},\ \bibinfo
  {pages} {3799} (\bibinfo {year} {2001})}\BibitemShut {NoStop}%
\bibitem [{\citenamefont {Hedayat}\ \emph {et~al.}(2019)\citenamefont
  {Hedayat}, \citenamefont {Sayers}, \citenamefont {Bugini}, \citenamefont
  {Dallera}, \citenamefont {Wolverson}, \citenamefont {Batten}, \citenamefont
  {Karbassi}, \citenamefont {Friedemann}, \citenamefont {Cerullo},
  \citenamefont {van Wezel}, \citenamefont {Clark}, \citenamefont {Carpene},\
  and\ \citenamefont {Da~Como}}]{Hedayat2019}%
  \BibitemOpen
  \bibfield  {author} {\bibinfo {author} {\bibfnamefont {H.}~\bibnamefont
  {Hedayat}}, \bibinfo {author} {\bibfnamefont {C.~J.}\ \bibnamefont {Sayers}},
  \bibinfo {author} {\bibfnamefont {D.}~\bibnamefont {Bugini}}, \bibinfo
  {author} {\bibfnamefont {C.}~\bibnamefont {Dallera}}, \bibinfo {author}
  {\bibfnamefont {D.}~\bibnamefont {Wolverson}}, \bibinfo {author}
  {\bibfnamefont {T.}~\bibnamefont {Batten}}, \bibinfo {author} {\bibfnamefont
  {S.}~\bibnamefont {Karbassi}}, \bibinfo {author} {\bibfnamefont
  {S.}~\bibnamefont {Friedemann}}, \bibinfo {author} {\bibfnamefont
  {G.}~\bibnamefont {Cerullo}}, \bibinfo {author} {\bibfnamefont
  {J.}~\bibnamefont {van Wezel}}, \bibinfo {author} {\bibfnamefont {S.~R.}\
  \bibnamefont {Clark}}, \bibinfo {author} {\bibfnamefont {E.}~\bibnamefont
  {Carpene}},\ and\ \bibinfo {author} {\bibfnamefont {E.}~\bibnamefont
  {Da~Como}},\ }\bibfield  {title} {\bibinfo {title} {{Excitonic and lattice
  contributions to the charge density wave in 1$T$-TiSe$_2$ revealed by a
  phonon bottleneck}},\ }\href
  {https://doi.org/10.1103/PhysRevResearch.1.023029} {\bibfield  {journal}
  {\bibinfo  {journal} {Phys. Rev. Research}\ }\textbf {\bibinfo {volume}
  {1}},\ \bibinfo {pages} {023029} (\bibinfo {year} {2019})}\BibitemShut
  {NoStop}%
\bibitem [{\citenamefont {Duan}\ \emph {et~al.}(2021)\citenamefont {Duan},
  \citenamefont {Cheng}, \citenamefont {Xia}, \citenamefont {Yang},
  \citenamefont {Xu}, \citenamefont {Qi}, \citenamefont {Huang}, \citenamefont
  {Tang}, \citenamefont {Guo}, \citenamefont {Luo}, \citenamefont {Qian},
  \citenamefont {Xiang}, \citenamefont {Zhang},\ and\ \citenamefont
  {Zhang}}]{Duan2021}%
  \BibitemOpen
  \bibfield  {author} {\bibinfo {author} {\bibfnamefont {S.}~\bibnamefont
  {Duan}}, \bibinfo {author} {\bibfnamefont {Y.}~\bibnamefont {Cheng}},
  \bibinfo {author} {\bibfnamefont {W.}~\bibnamefont {Xia}}, \bibinfo {author}
  {\bibfnamefont {Y.}~\bibnamefont {Yang}}, \bibinfo {author} {\bibfnamefont
  {C.}~\bibnamefont {Xu}}, \bibinfo {author} {\bibfnamefont {F.}~\bibnamefont
  {Qi}}, \bibinfo {author} {\bibfnamefont {C.}~\bibnamefont {Huang}}, \bibinfo
  {author} {\bibfnamefont {T.}~\bibnamefont {Tang}}, \bibinfo {author}
  {\bibfnamefont {Y.}~\bibnamefont {Guo}}, \bibinfo {author} {\bibfnamefont
  {W.}~\bibnamefont {Luo}}, \bibinfo {author} {\bibfnamefont {D.}~\bibnamefont
  {Qian}}, \bibinfo {author} {\bibfnamefont {D.}~\bibnamefont {Xiang}},
  \bibinfo {author} {\bibfnamefont {J.}~\bibnamefont {Zhang}},\ and\ \bibinfo
  {author} {\bibfnamefont {W.}~\bibnamefont {Zhang}},\ }\bibfield  {title}
  {\bibinfo {title} {{Optical manipulation of electronic dimensionality in a
  quantum material}},\ }\href {https://doi.org/10.1038/s41586-021-03643-8}
  {\bibfield  {journal} {\bibinfo  {journal} {Nature}\ }\textbf {\bibinfo
  {volume} {595}},\ \bibinfo {pages} {239} (\bibinfo {year}
  {2021})}\BibitemShut {NoStop}%
\bibitem [{\citenamefont {Hedayat}\ \emph {et~al.}(2021)\citenamefont
  {Hedayat}, \citenamefont {Sayers}, \citenamefont {Ceraso}, \citenamefont {van
  Wezel}, \citenamefont {Clark}, \citenamefont {Dallera}, \citenamefont
  {Cerullo}, \citenamefont {{Da Como}},\ and\ \citenamefont
  {Carpene}}]{Hedayat2021}%
  \BibitemOpen
  \bibfield  {author} {\bibinfo {author} {\bibfnamefont {H.}~\bibnamefont
  {Hedayat}}, \bibinfo {author} {\bibfnamefont {C.~J.}\ \bibnamefont {Sayers}},
  \bibinfo {author} {\bibfnamefont {A.}~\bibnamefont {Ceraso}}, \bibinfo
  {author} {\bibfnamefont {J.}~\bibnamefont {van Wezel}}, \bibinfo {author}
  {\bibfnamefont {S.~R.}\ \bibnamefont {Clark}}, \bibinfo {author}
  {\bibfnamefont {C.}~\bibnamefont {Dallera}}, \bibinfo {author} {\bibfnamefont
  {G.}~\bibnamefont {Cerullo}}, \bibinfo {author} {\bibfnamefont
  {E.}~\bibnamefont {{Da Como}}},\ and\ \bibinfo {author} {\bibfnamefont
  {E.}~\bibnamefont {Carpene}},\ }\bibfield  {title} {\bibinfo {title}
  {{Investigation of the non-equilibrium state of strongly correlated materials
  by complementary ultrafast spectroscopy techniques}},\ }\href
  {https://doi.org/10.1088/1367-2630/abe272} {\bibfield  {journal} {\bibinfo
  {journal} {New J. Phys.}\ }\textbf {\bibinfo {volume} {23}},\ \bibinfo
  {pages} {033025} (\bibinfo {year} {2021})}\BibitemShut {NoStop}%
\bibitem [{\citenamefont {Kaiser}(2017)}]{Kaiser2017}%
  \BibitemOpen
  \bibfield  {author} {\bibinfo {author} {\bibfnamefont {S.}~\bibnamefont
  {Kaiser}},\ }\bibfield  {title} {\bibinfo {title} {{Light-induced
  superconductivity in high-$T_c$ cuprates}},\ }\href
  {https://doi.org/10.1088/1402-4896/aa8201} {\bibfield  {journal} {\bibinfo
  {journal} {Phys. Scr.}\ }\textbf {\bibinfo {volume} {92}},\ \bibinfo {pages}
  {103001} (\bibinfo {year} {2017})}\BibitemShut {NoStop}%
\bibitem [{\citenamefont {Stahl}\ \emph {et~al.}(2020)\citenamefont {Stahl},
  \citenamefont {Kusch}, \citenamefont {Heinsch}, \citenamefont {Garbarino},
  \citenamefont {Kretzschmar}, \citenamefont {Hanff}, \citenamefont
  {Rossnagel}, \citenamefont {Geck},\ and\ \citenamefont
  {Ritschel}}]{Stahl2020}%
  \BibitemOpen
  \bibfield  {author} {\bibinfo {author} {\bibfnamefont {Q.}~\bibnamefont
  {Stahl}}, \bibinfo {author} {\bibfnamefont {M.}~\bibnamefont {Kusch}},
  \bibinfo {author} {\bibfnamefont {F.}~\bibnamefont {Heinsch}}, \bibinfo
  {author} {\bibfnamefont {G.}~\bibnamefont {Garbarino}}, \bibinfo {author}
  {\bibfnamefont {N.}~\bibnamefont {Kretzschmar}}, \bibinfo {author}
  {\bibfnamefont {K.}~\bibnamefont {Hanff}}, \bibinfo {author} {\bibfnamefont
  {K.}~\bibnamefont {Rossnagel}}, \bibinfo {author} {\bibfnamefont
  {J.}~\bibnamefont {Geck}},\ and\ \bibinfo {author} {\bibfnamefont
  {T.}~\bibnamefont {Ritschel}},\ }\bibfield  {title} {\bibinfo {title}
  {{Collapse of layer dimerization in the photo-induced hidden state of
  1$T$-TaS$_2$}},\ }\href {https://doi.org/10.1038/s41467-020-15079-1}
  {\bibfield  {journal} {\bibinfo  {journal} {Nat. Commun.}\ }\textbf {\bibinfo
  {volume} {11}},\ \bibinfo {pages} {1247} (\bibinfo {year}
  {2020})}\BibitemShut {NoStop}%
\bibitem [{\citenamefont {Qi}\ \emph {et~al.}(2020)\citenamefont {Qi},
  \citenamefont {Ma}, \citenamefont {Zhao}, \citenamefont {Cheng},
  \citenamefont {Jiang}, \citenamefont {Lu}, \citenamefont {Jiang},
  \citenamefont {Qian}, \citenamefont {Wang}, \citenamefont {Zhang},
  \citenamefont {Zhu}, \citenamefont {Zou}, \citenamefont {Wan}, \citenamefont
  {Xiang},\ and\ \citenamefont {Zhang}}]{Qi2020}%
  \BibitemOpen
  \bibfield  {author} {\bibinfo {author} {\bibfnamefont {F.}~\bibnamefont
  {Qi}}, \bibinfo {author} {\bibfnamefont {Z.}~\bibnamefont {Ma}}, \bibinfo
  {author} {\bibfnamefont {L.}~\bibnamefont {Zhao}}, \bibinfo {author}
  {\bibfnamefont {Y.}~\bibnamefont {Cheng}}, \bibinfo {author} {\bibfnamefont
  {W.}~\bibnamefont {Jiang}}, \bibinfo {author} {\bibfnamefont
  {C.}~\bibnamefont {Lu}}, \bibinfo {author} {\bibfnamefont {T.}~\bibnamefont
  {Jiang}}, \bibinfo {author} {\bibfnamefont {D.}~\bibnamefont {Qian}},
  \bibinfo {author} {\bibfnamefont {Z.}~\bibnamefont {Wang}}, \bibinfo {author}
  {\bibfnamefont {W.}~\bibnamefont {Zhang}}, \bibinfo {author} {\bibfnamefont
  {P.}~\bibnamefont {Zhu}}, \bibinfo {author} {\bibfnamefont {X.}~\bibnamefont
  {Zou}}, \bibinfo {author} {\bibfnamefont {W.}~\bibnamefont {Wan}}, \bibinfo
  {author} {\bibfnamefont {D.}~\bibnamefont {Xiang}},\ and\ \bibinfo {author}
  {\bibfnamefont {J.}~\bibnamefont {Zhang}},\ }\bibfield  {title} {\bibinfo
  {title} {{Breaking 50 femtosecond resolution barrier in MeV ultrafast
  electron diffraction with a double bend achromat compressor}},\ }\href
  {https://doi.org/10.1103/PhysRevLett.124.134803} {\bibfield  {journal}
  {\bibinfo  {journal} {Phys. Rev. Lett.}\ }\textbf {\bibinfo {volume} {124}},\
  \bibinfo {pages} {134803} (\bibinfo {year} {2020})}\BibitemShut {NoStop}%
\end{thebibliography}
\end{document}